\documentclass[twocolumn,superscriptaddress,floatfix,preprintnumbers,prd,nofootinbib,hyperref]{revtex4-2}

\usepackage{graphicx,color,amsmath,amssymb}
\usepackage{slashed}
\usepackage{soul}
\usepackage{amsmath}
\usepackage{braket}
\usepackage[utf8]{inputenc}
\usepackage{aas_macros}
\usepackage{float}

\newcommand{\nocontentsline}[3]{}
\newcommand{\toclesslab}[3]{\bgroup\let\addcontentsline=\nocontentsline#1{#2\label{#3}}\egroup}
\newcommand{\tocless}[2]{\bgroup\let\addcontentsline=\nocontentsline#1{#2}\egroup}
\newcommand{\epar}{E_\parallel}

\newcommand{\rpc}{r_{\text{pc}}}

\newcommand{\gagg}{g_{a\gamma\gamma}}

\usepackage[export]{adjustbox}
\usepackage{tikz}
\usepackage{tkz-euclide}
\usetikzlibrary{decorations.pathmorphing}	
\tikzset{
    v/.style={decorate, decoration={snake, segment length=3mm, amplitude=0.75mm}, draw},
    f/.style={draw=black, postaction={decorate},
        decoration={markings,mark=at position .6 with {\arrow[very thick]{latex}}}},
    fb/.style={draw=black, postaction={decorate},
        decoration={markings,mark=at position .4 with {\arrowreversed[very thick]{latex}}}},
    fnar/.style={draw=black},
    g/.style={decorate, draw=black,
        decoration={coil,amplitude=3pt, segment length=3.5pt}},
    s/.style={dashed,draw=black, postaction={decorate},
        decoration={markings,mark=at position .55 with {\arrow[very thick]{latex}}}},
    sb/.style={dashed,draw=black, postaction={decorate},
        decoration={markings,mark=at position .55 with {\arrowreversed[draw=black,very thick]{latex}}}},
    snar/.style={dashed,draw=black,line width =1.25pt},
}

\usepackage{hyperref} 
\usepackage{amsmath}
\hypersetup{
    colorlinks=true,
    citecolor=[rgb]{.1, .7, .6},
    linkcolor=[rgb]{.2, .55, .95},
    filecolor=magenta,
    urlcolor=[rgb]{.1, .7, .6},
}

\newcommand{\km}{{\, {\rm km}}}

\newcommand{\eV}{{\, {\rm eV}}}

\newcommand{\GeV}{{\, {\rm GeV}}}

\newcommand{\MHz}{{\, {\rm MHz}}}

\newcommand{\ie}{{\it i.e.~}}  \newcommand{\eg}{{\it e.g.~}}
\definecolor{mypurple}{RGB}{164,64,214}

\definecolor{darkgreen}{rgb}{0.0, 0.5, 0.0}

\begin{document}

\title{Axion Clouds around Neutron Stars}
\author{Dion Noordhuis}
\email{d.noordhuis@uva.nl}
\affiliation{GRAPPA Institute, Institute for Theoretical Physics Amsterdam and Delta
Institute for Theoretical Physics,
University of Amsterdam, Science Park 904, 1098 XH Amsterdam, The Netherlands}
\author{Anirudh Prabhu}
\email{prabhu@princeton.edu}
\affiliation{Princeton Center for Theoretical Science, Princeton University, Princeton, New Jersey 08544, USA}
\author{Christoph Weniger}
\email{c.weniger@uva.nl}
\affiliation{GRAPPA Institute, Institute for Theoretical Physics Amsterdam and Delta
Institute for Theoretical Physics,
University of Amsterdam, Science Park 904, 1098 XH Amsterdam, The Netherlands}
\author{Samuel J. Witte}
\email{samuel.witte@physics.ox.ac.uk}
\affiliation{GRAPPA Institute, Institute for Theoretical Physics Amsterdam and Delta
Institute for Theoretical Physics,
University of Amsterdam, Science Park 904, 1098 XH Amsterdam, The Netherlands}
\affiliation{Departament de F\'{i}sica Qu\`{a}ntica i Astrof\'{i}sica and Institut de Ciencies del Cosmos (ICCUB) ,
Universitat de Barcelona, Diagonal 647, E-08028 Barcelona, Spain}
\affiliation{Rudolf Peierls Centre for Theoretical Physics, University of Oxford, Parks Road, Oxford OX1 3PU, United Kingdom}


\begin{abstract}
Recent work has shown that axions can be efficiently produced via non-stationary pair plasma discharges in the polar cap region of pulsars. Here, we point out that for axion masses $10^{-9} \, {\rm eV} \lesssim m_a \lesssim 10^{-4} \, \rm eV$, a sizable fraction of the sourced axion population will be gravitationally confined to the neutron star. These axions accumulate over astrophysical timescales, thereby forming a dense `axion cloud' around the star. We argue that the existence of such a cloud, with densities reaching and potentially exceeding $\mathcal{O}(10^{22}) \, {\rm GeV \, cm^{-3}}$, is a generic expectation across a wide range of parameter space. For axion masses $m_a \gtrsim 10^{-7} \eV$, energy is primarily radiated from the axion cloud via resonant axion-photon mixing, generating a number of distinctive signatures that include: a sharp line in the radio spectrum of each pulsar (located at the axion mass, and with an order percent-level width), and transient events arising from the reconfiguration of charge densities in the magnetosphere. While a deeper understanding of the systematic uncertainties in these systems is required, our current estimates suggest that existing radio telescopes could improve sensitivity to the axion-photon coupling by more than an order of magnitude.
\end{abstract}

\maketitle

\section{Introduction}
A generic prediction of several extensions of the Standard Model of particle physics is the existence of light pseudoscalars known as axions. Perhaps the best-known example is the `QCD axion', a pseudo-Goldstone boson arising in the leading solution to the Strong-CP problem (this is the question of why charge-parity symmetry is preserved in quantum chromodynamics)~\cite{PQ1, PQ2, WeinbergAxion, WilczekAxion}. Light axions are also a standard feature appearing from the compactification of extra dimensions in well-motivated extensions of the Standard Model (such as \eg string theory~\cite{Arvanitaki:2009fg, Witten:1984dg, Cicoli:2012sz, Conlon:2006tq, Svrcek:2006yi}), and are among the leading candidates to explain the `missing matter' in the Universe, \ie dark matter (see \eg~\cite{Adams:2022pbo} for a recent review).

One of the novel and more compelling proposals to indirectly search for axions involves looking for radio signals coming from the magnetospheres of neutron stars. The large magnetic fields and ambient plasma found in these environments can dramatically enhance the interaction rate between axions and electromagnetism, producing an array of distinctive signatures. These include radio spectral lines arising from the resonant conversion of axion dark matter~\cite{Pshirkov:2007st,Huang:2018lxq,Hook:2018iia,Leroy:2019ghm,Safdi:2018oeu, Battye:2019aco,Foster:2020pgt, Witte2021,millar2021axionphotonUPDATED, battye2021robust,Foster:2022fxn, Witte:2022cjj,Battye:2023oac}, broadband radio emission generated from axions sourced in the polar caps of pulsars~\cite{Prabhu:2021zve,Noordhuis:2022ljw}, and transient radio bursts~\cite{Iwazaki:2014wka,Bai:2017feq,Dietrich:2018jov,Prabhu:2020yif,Edwards:2020afl,Buckley:2020fmh,Nurmi:2021xds,Bai:2021nrs,Witte:2022cjj,Prabhu:2023cgb}. 


In this article, we show that the mere existence of an axion with a mass in the range $10^{-9} \, {\rm eV} \lesssim m_a \lesssim 10^{-4} \, {\rm eV}$ leads to the generic prediction that all neutron stars are surrounded by extremely dense clouds of axions -- this statement does not require axions to contribute to the dark matter density of the Universe, and remains valid even for very feebly coupled axions, including the QCD axion. This is a direct consequence of the fact that iterative bursts of $e^\pm$ pair production taking place in the polar caps of active neutron stars induce a quasiperiodic dynamical screening of $\vec{E} \cdot \vec{B}$, which enters as a source term in the axion's equation of motion (here $\vec{E}$ and $\vec{B}$ are the electric and magnetic field supported by the neutron star). Ref.~\cite{Noordhuis:2022ljw} recently investigated the observable consequences arising from the high-energy part of the sourced axion population (\ie the part which escapes the magnetosphere), showing that these axions can give rise to a substantial broadband radio flux. However, not all of the produced axions escape. For axion masses roughly in the megahertz to gigahertz range, a large fraction of injected energy will go directly into axions that are instead gravitationally confined to the neutron star. Owing to their feeble interactions, these particles cannot efficiently dissipate energy, and will therefore accumulate on astrophysical timescales. Importantly, the enormous densities realized in these environments allow one to overcome the traditional challenges associated with detecting feebly interacting particles; as such, the  existence of dense axion fields around neutron stars will not only have a profound impact on existing indirect axion searches (such as those in~\cite{battye2021robust,Foster:2022fxn, Noordhuis:2022ljw,Battye:2023oac}), but will also open a new, cross-disciplinary regime of axion phenomenology that carries enormous discovery potential.  


The goal of this paper is to explore the formation, the properties, and the evolution of axion clouds around neutron stars, setting the stage for future studies of the phenomenological implications of these systems. Axion production near the polar cap regions of rotating neutron stars was previously studied in~\cite{Prabhu:2021zve, Noordhuis:2022ljw}. This paper builds upon these works, investigating for the first time the properties of axions that remain gravitationally bound to the star. Understanding the long-term evolution and implications of these axions requires detailed modeling of axion production, axion phase space evolution, and neutron star evolution, on timescales ranging from microseconds to megayears and spatial scales ranging from centimeters to hundreds of kilometers. 
We achieve this by simulating a small population of neutron stars from birth to death, tracing the change in the axion production rate due to magnetorotational spin-down (by looking at the temporal evolution of the axion phase space density around the neutron star), and determining the extent to which the energy stored in the axion cloud can be dissipated over the course of a neutron star's lifetime. We find that for typical neutron stars in our sample, the characteristic density of axions near the surface of the star can exceed the local dark matter density by more than 20 orders of magnitude (with large densities achieved even for axion-photon couplings as low as $\gagg \sim M_{\rm pl}^{-1}$) over a sizable fraction of the neutron star's lifetime. Finally, we identify a number of unavoidable phenomenological implications of axion clouds, including the emission of a sharp spectral line at radio frequencies and a transient decaying radio burst. We demonstrate that observations using current radio telescopes offer enormous discovery potential, including for the QCD axion.
We conclude by emphasizing that the generic existence of such dense axion configurations around neutron stars opens up numerous phenomenological avenues that will be the subject of future work.

This article is organized as follows. We begin in Sec.~\ref{sec:formation} by discussing the fundamental physics governing the formation and evolution of axion clouds. This includes a thorough discussion of various mechanisms capable of dissipating the stored energy; here, we show that the dominant energy dissipation mechanism is typically resonant axion-photon mixing. One inevitable consequence that arises from this mixing is the production of radio emission -- this is the focus of Sec.~\ref{sec:radio}. In this section we identify two unique signatures arising from the existence of axion clouds\footnote{A companion paper, which includes one of the authors of this paper, also illustrates how at low axion masses, below those studied here, the axion cloud can dissipate energy by driving radiative energy losses in the plasma itself. In this regime, the axion cloud can backreact on the electrodynamics in the polar cap, imprinting a periodic nulling on a pulsar's radio emission~\cite{PulsarNulling}.}: a sharp kinematic feature in the radio band appearing at the axion mass, and a transient radio burst which serves to dissipate the axion cloud late in a neutron star's lifetime. We furthermore estimate projected sensitivities of existing radio telescopes to both signatures in Sec.~\ref{sec:sensitivity}. We provide concluding remarks in Sec.~\ref{sec:conclusions}.

\section{Axion clouds}\label{sec:formation}
The first model for a plasma-filled magnetosphere was introduced more than 50 years ago, when Ref.~\cite{GoldreichJulian1969} noted that the parallel component of the induced electric field, $\epar \equiv (\vec{E}\cdot\vec{B}) / |\vec{B}| \neq 0$, can easily overpower gravity, naturally leading to the extraction of charges directly from the upper layers of a neutron star's surface. The extracted current drives electric fields which attempt to screen $\epar$, eventually pushing the system toward a steady-state configuration in which $\epar = 0$ everywhere~\cite{GoldreichJulian1969}. This solution, known as the Goldreich-Julian (GJ) model, requires the plasma to corotate with the neutron star, a feat which is impossible at sufficiently large radial distances. 

As a result, small localized regions known as vacuum gaps, with $\epar \neq 0$, are expected to appear in order to supplement the plasma deficit at large radii~\cite{RudermanSutherland1975}. As mentioned above, however, the existence of a large $\epar$ component is unstable. The parallel electric field will extract and accelerate primary particles from the neutron star surface -- these particles will subsequently ignite $e^\pm$ pair cascades, generating a dense plasma which temporarily screens $\epar$ before flowing away from the neutron star. Recent years have shown significant progress in simulating this dynamical discharge process taking place in the polar cap gaps (\ie the vacuum gaps situated just above the magnetic poles of a neutron star), indicating that it may be the source of the coherent radio emission observed from pulsars~\cite{Spitkovsky2020, Cruz2021}.

\begin{figure*}
    \centering
    \includegraphics[width=\textwidth]{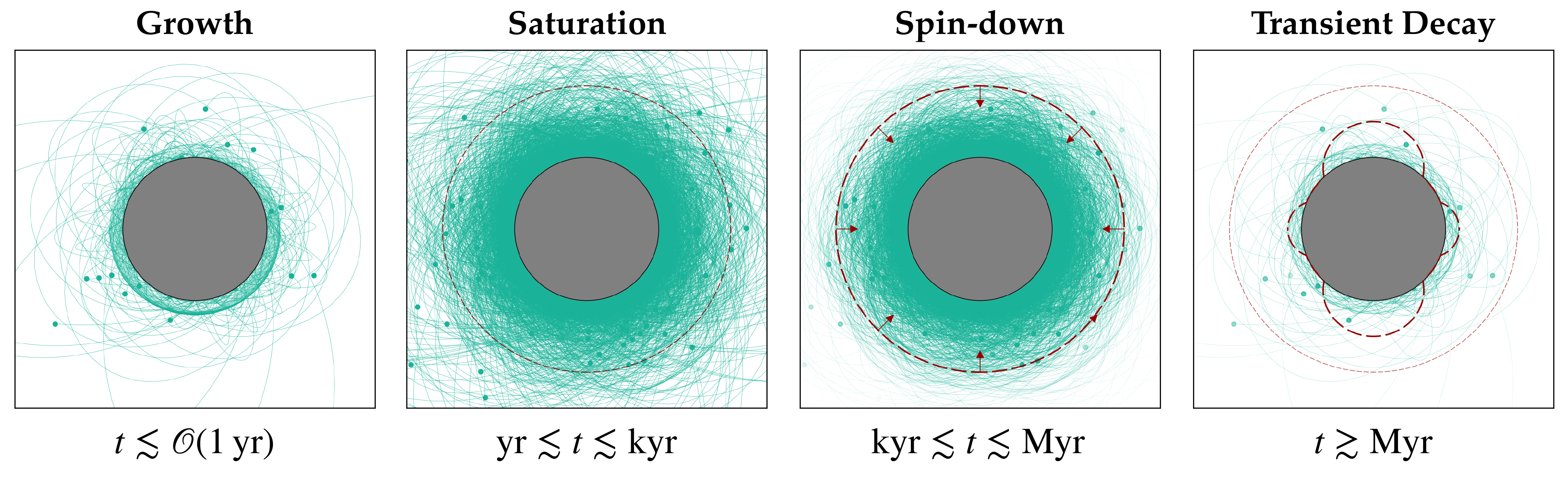}
    \caption{Overview of the four stages characterizing the formation and evolution of axion clouds around neutron stars. {\bf Growth} (left): typically occurring on sub-year timescales, the axion cloud grows in an unimpeded manner. {\bf Saturation} (center-left): occurring on $\mathcal{O}({\rm yr})$ to $\mathcal{O}({\rm kyr})$ timescales, the outer part of the density profile saturates as a result of resonant axion-photon conversion (taking place near the conversion surface shown as a dashed red line). The inner part of the profile is unaffected by the resonant transitions, and continues to grow. {\bf Spin-down} (center-right): occurring on $\mathcal{O}({\rm kyr})$ to $\mathcal{O}({\rm Myr})$ timescales, the neutron star undergoes magnetorotational spin-down, causing the production rate of axions to decrease and shifting the resonant conversion surface to smaller radii. {\bf Transient Decay} (right): late in the neutron star's lifetime axion production ceases and the magnetosphere relaxes to a fully charge-separated state. This process opens resonant transitions at all radii, and causes the axion cloud to fully dissipate its energy. }
    \label{fig:overview}
\end{figure*}

It was recently demonstrated that the dynamical screening of $\epar$ in the polar caps can also efficiently source a local population of axions~\cite{Prabhu:2021zve,Noordhuis:2022ljw} -- this is a consequence of the fact that $\vec{E} \cdot \vec{B}$ enters as a source term in the axion's equation of motion. The spectrum of emitted axions is determined by the induced oscillations that arise from the screening process, and is expected to roughly span frequencies in the megahertz to gigahertz range (consistent with the spectrum of observed radio emission). Relativistic axions produced in this way can escape the gravitational pull of the neutron star -- during their traversal of the magnetosphere these axions can, on occasion, resonantly convert to low-energy radio photons. Ref.~\cite{Noordhuis:2022ljw} used a combination of semi-analytic modeling, numerical simulations, and geometric ray-tracing to compute the spectra of escaping axions and photons sourced in the magnetospheres of nearby pulsars. By comparing the predicted and observed radio flux from this pulsar population, Ref.~\cite{Noordhuis:2022ljw} was able to set stringent constraints on the axion-photon coupling across a wide range of axion masses.

In this paper, we focus instead on the implications arising from non-relativistic axions sourced in the polar caps, which can comprise a sizable fraction of the total axion population if the axion mass is roughly in the range $10^{-9} \, {\rm eV} \lesssim m_a \lesssim 10^{-4} \, {\rm eV}$. These non-relativistic axions are gravitationally bound to the neutron star, and will therefore accumulate over time, resulting in the formation of a dense axion cloud\footnote{Axion clouds share some similarities with both superradiant clouds of bosonic particles around black holes (see \eg~\cite{Arvanitaki:2009fg, Arvanitaki:2010sy, Brito:2015oca, Baryakhtar:2017ngi, Baryakhtar_2021}), and basins of light bosonic particles around stellar objects (see \eg~\cite{VanTilburg2021, Lasenby2021}). Black hole superradiance and the non-thermal axion production mechanism studied here operate by extracting the rotational energy from dense astrophysical objects and using this energy to produce gravitationally bound bosons. Superradiant clouds, however, grow exponentially (while the growth of neutron star axion clouds is at most linear) and this phenomenon is only efficient when bosons are sufficiently light ($m_a \lesssim 10^{-11} \, \rm eV$ for astrophysical black holes). Stellar basins, on the other hand, are produced via thermal processes, and thus large densities can only be achieved for masses close to the thermal temperature of the emitting astrophysical body. These studies are qualitatively similar to the analysis presented here, but rely on volumetric thermal emission in the Sun. In contrast, we study non-thermal emission arising from pulsar polar caps; the case of thermal emission from neutron stars is not expected to produce large densities as it necessitates much larger axion masses in order to be effective, and these axions are efficiently absorbed inside of the neutron star.}. In this section we study the evolution of these clouds over the lifetime of the neutron star, showing that an axion cloud proceeds through four distinct phases -- these phases are highlighted in Fig.~\ref{fig:overview}.

\subsection{General evolution}
Axions generically couple to electromagnetism via the Lagrangian term $\mathcal{L} \supset - \frac{1}{4} \gagg \, a \, F_{\mu\nu} \, \tilde{F}^{\mu\nu}$, where $F$ and $\tilde{F}$ are the electromagnetic field strength tensor and its dual, $a$ is the axion field, and $g_{a\gamma\gamma}$ is the coupling strength. The resulting axion equation of motion is given by
\begin{equation}
    \left(\Box + m_a^2 \right) a(x) = \gagg \, ( \vec{E} \cdot \vec{B})(x) \, ,
\end{equation}
where $\vec{E} \cdot \vec{B}$ clearly appears as an axion source term. Following Ref.~\cite{Noordhuis:2022ljw}, one can express the differential production rate of axions from spacetime variations in the source term via
\begin{equation}\label{eqn:diffrate}
    \frac{d \dot{N}_a}{d^3 k}  = \frac{\left| \tilde{\mathcal{S}}(\vec{k}) \right|^2}{2 (2\pi)^3 \omega_a(\vec{k}) T} \, .
\end{equation}
Here $\omega_a(\vec{k})$ is the energy of mode $\vec{k}$, $T$ is the characteristic period of the gap discharge process, and $\tilde{\mathcal{S}}(\vec{k})$ is the Fourier transform of the source term, \ie
\begin{equation}\label{eq:fftj}
    \tilde{\mathcal{S}}(\vec k) = \displaystyle \gagg \int d^4 x \, e^{i k \cdot x} (\vec{E} \cdot \vec{B})(x) \, ,
\end{equation}
where the temporal integral is performed over the gap period and the spatial integral over the volume of the gap. The geometry of the gap is approximated as a cylinder of radius $\rpc = R_{\rm NS} \sqrt{\Omega_{\rm NS} R_{\rm NS}}$, where $R_{\rm NS}$ and $\Omega_{\rm NS}$ are the neutron star radius and rotational velocity respectively, and height $h_{\rm gap}$. The explicit expression for $h_{\rm gap}$ is given in~\cite{Noordhuis:2022ljw, PulsarNulling}. From Eq.~\ref{eq:fftj}, one can see that the production rate and spectrum of axions produced during the screening of the vacuum gap is fully resolved by the spacetime evolution of $\epar$, which is non-trivial due to pair production processes  (examples of the axion spectrum can be found in Fig. S4 of the Supplemental Material of~\cite{Noordhuis:2022ljw}).

\begin{table*}
    \centering
    \begin{tabular}{|c|c|c|c|c|c|c|}
    \hline
    NS & $P_{\rm birth}$ (s) & $B_{0, \rm birth}$ (G) & $\chi_{\rm birth}$ (deg) & $P_{\rm death}$ (s) & $B_{0, \rm death}$ (G) & $\chi_{\rm death}$ (deg) \\ \hline \hline
    1 & 0.186 & 2.88E13 & 48 & 3.40 & 3.94E12 & 5 \\ \hline
    2 & 0.168 & 1.39E14 & 23 & 10.8 & 3.99E13 & 0.41 \\ \hline
    3 & 0.094 & 1.73E13 & 46 & 2.16 & 1.59E12 & 3.7 \\ \hline
    4 & 0.429 & 2.24E12 & 62 & 0.587 & 1.17E11 & 58 \\ \hline
    5 & 0.158 & 3.76E13 & 9.7 & 4.18 & 5.95E12 & 0.37 \\ \hline
    6 & 0.533 & 5.89E12 & 80 & 1.21 & 4.97E11 & 75 \\ \hline
    7 & 0.486 & 1.49E13 & 74 & 2.51 & 2.14E12 & 55 \\ \hline
    8 & 0.990 & 2.00E13 & 14 & 2.65 & 2.38E12 & 5.5 \\ \hline
    9 & 0.700 & 1.04E12 & 51 & 0.722 & 1.78E11 & 50 \\ \hline
    10 & 0.054 & 1.00E13 & 86 & 1.76 & 1.06E12 & 65 \\ \hline
    \end{tabular}
    \caption{Listed properties of the mock neutron stars used in this work. The columns denote neutron star number, period at birth (in seconds), surface magnetic field strength at birth (in Gauss), misalignment angle at birth (in degrees), period at death (in seconds), surface magnetic field strength at death (in Gauss), and misalignment angle at death (in degrees). }
    \label{tab:pulsars}
\end{table*}

In order to make general statements about the properties of axion clouds, we simulate the evolution of ten different neutron stars over the course of their lifetime. Following Refs.~\cite{popov2010population,gullon2014population}, we assume the initial period ($P$ in seconds), magnetic field ($B_0$ in Gauss), and misalignment angle ($\chi$ in degrees) are uncorrelated at birth, and characterized by the following distributions
\begin{eqnarray}\label{eq:spindown}
p(P) &=& \frac{1}{\sqrt{2\pi \sigma_p^2} }\, e^{-(P - \mu_p)^2/(2 \sigma_p^2)} \, , \\
p(B_0) &=& \frac{1}{\sqrt{2\pi \sigma_B^2} }\, e^{-(\log_{10}(B_0) - \mu_B)^2/(2 \sigma_B^2)} \, , \\
p(\chi) &=& \sin\chi / 2 \, . \label{eq:spindown_chi}
\end{eqnarray}
Here, the means and standard deviations have been obtained by performing fits to the observed pulsar population, and are given by $\mu_p=0.22$, $\sigma_p = 0.42$\footnote{Note that the distribution of the period is restricted to the positive definite domain, and the misalignment angle $\chi$ is restricted to be between $0$ and $90$ degrees (pulsar evolution proceeds identically regardless of whether the projection of the magnetic axis onto the rotational axis is positive or negative).}, $\mu_B = 13.2$, and $\sigma_B = 0.62$. The initial conditions for the ten neutron stars in our sample are obtained via random draws from Eqs.~\ref{eq:spindown}--\ref{eq:spindown_chi}; the properties of each neutron star at birth are presented in the first columns of Table~\ref{tab:pulsars}.

For active neutron stars, the rate of spin-down is determined by a combination of dipole radiation and plasma effects (see \eg~\cite{Spitkovsky2006,Philippov2014}). The evolution of the rotational period and misalignment angle are given by
\begin{eqnarray}
\dot{P} &=& \beta \frac{B_0^2}{P} \, (\kappa_0+ \kappa_1\sin^2\chi) \, , \label{eqn:pevol} \\
\dot{\chi} &=& -\beta \, \kappa_2 \frac{B_0^2}{P^2} \, \sin\chi\cos\chi \, , \label{eqn:chievol}
\end{eqnarray}
where $\kappa_0 \sim \kappa_1 \sim \kappa_2 \sim 1$ and $\beta = \pi^2 R_{\rm NS}^6 / I_{\rm NS} \sim 6 \times 10^{-40} \, {\rm G^{-2} \, s}$, with $I_{\rm NS}$ being the neutron star moment of inertia.

Detailed modeling of magnetic field evolution in neutron stars is an active area of research that involves detailed simulations of internal neutron star conductivity (see~\cite{Bransgrove2018, Igoshev2021, Vigano2021} for recent progress). The main processes governing the evolution of neutron star magnetic fields are Ohmic dissipation, Hall drift, and ambipolar diffusion. Ohmic dissipation simply describes the energy losses that arise from having a finite conductivity internal to the neutron star. The rate of Ohmic dissipation can be enhanced by Hall drift, which allows for magnetic flux to be transported from inner regions of the star, where the rate of energy dissipation is slow, to the outer crust, where the rate is considerably higher. Ambipolar diffusion arises due to relative motion between charged particles and neutrons, leading to a drag that dissipates energy -- this dissipation mechanism is typically efficient only early in the lifetime of neutron stars with extremely large magnetic fields. Here, we make the simplifying assumption that magnetic field evolution is dominated by Ohmic dissipation, which drives exponential decay of the magnetic field on a timescale $\tau_{\rm Ohm}$, \ie
\begin{eqnarray}
\dot{B_0} &=&  - B_0 / \tau_{\rm Ohm} \label{eqn:bevol} \, .
\end{eqnarray}
We take $\tau_{\rm Ohm}$ to be $1 \, \rm Myr$~\cite{Vigano:2013lea,gullon2014population}, which is expected to lie on the very conservative\footnote{Note that small Ohmic decay times are `conservative' in that, under the assumption of a fixed neutron star formation rate, they suppress the size of the active pulsar population.} end of the spectrum as the simulations of~\cite{Vigano:2013lea,gullon2014population} assume that the currents supporting the magnetic field are confined to the crust -- should the currents also penetrate the core, as may be expected, the Ohmic decay timescale could increase by orders of magnitude. Our treatment neglects the impact of Hall drift and ambipolar diffusion, which can induce small corrections to high-magnetic field neutron stars on shorter timescales. Properly including such contributions, however, requires more refined modeling and a better understanding of the behavior of currents in the stellar interior. 

The properties of the ten neutron stars considered in this work, at both birth and death, are displayed in Table~\ref{tab:pulsars}\footnote{Notice that the magnetic field of one of our neutron stars exceeds the Schwinger field strength $B_Q \simeq 4.4 \times 10^{13} \, \rm G$. In this limit, the gap dynamics and charge configuration may deviate from that of standard pulsars. In what follows, we treat this neutron star with the same dynamics as the others, but we avoid making any statements that rely on the properties of this outlier.}. We define the point of death as the time at which the neutron star is no longer able to produce radio emission -- this occurs when the rotational period falls below $P \simeq 1.7 \, {\rm s} \,  \sqrt{B_0 / 10^{12} \, {\rm G}}$~\cite{Johnston2017}. Once a neutron star crosses this point we also assume that it is no longer able to generate axions. For each neutron star in our sample, we utilize the semi-analytic model of~\cite{Noordhuis:2022ljw} to predict the axion spectrum at birth, $10 \, \rm yrs$, $100 \, \rm yrs$, $1 \, \rm kyr$, $10 \, \rm kyrs$, $100 \, \rm kyrs$, and at death.  These spectra allow for the smooth interpolation of the production rate of axions throughout the lifetime of the neutron star. 

The time evolution of the axion spectra can subsequently be used to reconstruct the density profile of axions around the neutron stars at any given time. In order to achieve this we first note that Eq.~\ref{eqn:diffrate} describes the number of axions produced with a given $k$-mode across the entirety of the gap -- if the gap is sufficiently small (such that the gradient of the gravitational potential across the gap can be neglected), one can assume all axions in a narrow range of $k$-modes follow approximately the same trajectory. Adopting this approximation, we take $\sim 10^4$ logarithmically spaced bins over momenta $k < k_{\rm esc}$ (where $ k_{\rm esc}$ is the escape momentum), and trace the evolution of these bound trajectories over timescales on the order of $\tau_0 \sim \mathcal{O}(1 - 10^{3})$ seconds (corresponding to $10^{3} - 10^{4}$ neutron star crossing times). Considering all trajectories, we identify the points at which they cross a given shell of radius $r$, and average the energy density of axions contained in the trajectories over this shell -- this procedure yields a radial density profile at $\tau_0$ given by
\begin{eqnarray}\label{eq:rho_remake}
    \rho(r, \tau_0) &=& \frac{1}{A} \int dA \, n_a(r) \, \omega_a(r) \nonumber \\
    &\sim&  \frac{1}{4\pi r^2} \sum_{k_j, x} \frac{\omega_{j,x} \, \dot{N}_{j,x}}{v_{j,x} \cos\theta_{j,x}} \, .
\end{eqnarray}
Here we have used the fact that one can express the number density $n_a$ in terms of the axion production rate $\dot{N_a}$, \ie $n_a \equiv dN_a / d^3x = dN_a / (v_a \cos\theta \, dA \, dt ) $ (where $dA$ is the area element with a surface normal oriented at an angle $\theta$ with respect to the axion velocity $v_a$). The summation runs over the trajectories $k_j$ and the crossings $x$ of each trajectory with the radial surface located at $r$. 

In the event that axions can be treated as fully non-interacting, the density profile is expected to grow linearly as long as the properties of the neutron star remain constant (\ie on timescales shorter than the spin-down and magnetic field decay timescales). This is conservatively expected to last on the order of $\tau_{\rm linear} \sim \mathcal{O}({\rm kyr})$. On longer timescales, magnetorotational spin-down decreases axion production, leading to an eventual saturation of the growth. We incorporate the effect of spin-down into the evolution of the density profile at a given time $\tau$ by rescaling Eq.~\ref{eq:rho_remake} by a factor of $\tau/\tau_0$, and redefining $\dot{N}_{j,x}$ with an effective production rate $\dot{N}_{j,x}^{\rm eff}$, which is obtained at each $\tau$ by interpolating between the axion production rates of the trajectories as computed over the lifetime of the neutron star. As we discuss below, there moreover exist various mechanisms which can dissipate energy or disrupt the production of axions. As a result, for large axion couplings the assumption of linear growth on short timescales may not hold, and the growth of the cloud may be saturated long before spin-down substantially abates axion production. 

\subsection{Energy dissipation mechanisms}
In order to properly follow the long-term evolution of the axion cloud, one must understand the extent to which the feeble interactions of the axion can dissipate energy, or backreact on the production process itself. Here, we investigate a number of mechanisms which in this way could alter the linear growth of the axion density. These include: absorption by nuclei internal to the neutron star, axion self-interactions, and electromagnetic interactions. Our calculations suggest that the dominant mechanisms are typically expected to be electromagnetic.

\subsubsection{Absorption}
Axions have a dimension-five coupling to nucleons via the Lagrangian term
\begin{eqnarray}
    \mathcal{L} \supset \frac{1}{2} g_{aN} \, \partial_\mu a \overline{N} \, \gamma^\mu \gamma_5 \, N \, ,
\end{eqnarray}
which can induce inverse bremsstrahlung absorption (or bremsstrahlung emission) of axions as they traverse the nuclear matter interior to the neutron star\footnote{Other absorption channels are also possible, but always subdominant to axion-nucleon bremsstrahlung~\cite{Iwamoto:1984, Brinkmann:1988, Buschmann:2022}.}. Here, we show that for typical axion-nucleon couplings $g_{aN}$ this absorption process is not expected to significantly impede the evolution of the cloud.

We start by noting that the evolution of a single axion trajectory with phase space $f_a$ can be described by the one-dimensional Boltzmann equation (in flat space)
\begin{equation}
    (\partial_t + v_a \partial_x)f_a = \Gamma_E (1+f_a) - \Gamma_A f_a = \Gamma_E - \Gamma_{A*} f_a \, ,
\end{equation}
where $\Gamma_E$ and $\Gamma_A$ are the emission and absorption rate in vacuum, respectively. In the final expression we have defined the effective absorption rate $\Gamma_{A*} \equiv \Gamma_A - \Gamma_E$, which accounts for the fact that a high occupation number axion background can also stimulate the bremsstrahlung emission of new axions. If the nuclear matter is in local thermal equilibrium, detailed balance implies that locally $\Gamma_E = e^{-\omega_a/T_{\rm NS}} \Gamma_A$, with $\omega_a$ being the axion energy and $T_{\rm NS}$ being the neutron star temperature (see \eg~\cite{raffelt1996stars}). Since $\omega_a \ll T_{\rm NS}$ for all axions in the cloud, it follows that $\Gamma_{A*} \approx (\omega_a / T_{\rm NS}) \Gamma_{A} \ll \Gamma_{A}$, and thus the absorption rate tends to be heavily suppressed in these systems. 

In order to provide a quantitative estimate of the absorption timescale, we use the Fermi surface approximation to compute the axion-nucleon bremsstrahlung mean free path. Details can, for example, be found in~\cite{Iwamoto:2001}, where the mean free path is given in Eq.~3.14. Assuming the mean free path is constant through the neutron star, the fraction of axions absorbed via a single pass is given by
\begin{equation}\label{eq:fabsorb}
    f_{\rm abs} = \frac{N_{\rm ab}}{N_0} = \left(1 - e^{-x / l_a^{\rm eff}}\right) \, ,
\end{equation}
where $x$ is the distance traveled inside of the star and $l_a^{\rm eff} = l_a / (\omega_a / T_{\rm NS})$ is the effective mean free path (including the factor due to stimulated emission). By tracing axion trajectories over many crossings and applying Eq.~\ref{eq:fabsorb} to each, we can determine the typical timescale on which axion absorption takes place -- we define this timescale to be $\tau_{90}$, which is the time after production when $90\%$ of the axions on a given trajectory will have been absorbed.

In computing the absorption cross section we adopt a characteristic value for the internal temperature of $T_{\rm NS} = 10^6 \, \rm K$ and the nucleon Fermi momenta of $p_{F_N} \sim 300 \, \rm MeV$. Furthermore using an axion-nucleon coupling of $g_{aN} = 10^{-10} \, \rm GeV^{-1}$, we find that low-energy axions (whose maximal radial distance from the neutron star is $r_{\rm max} \leq 20 \, \rm km$) have $ \tau_{90} \sim 100 \, \rm yrs - 1 \, \rm kyr$. We determined empirically that this timescale grows roughly $\propto r_{\rm max}^{1.4}$ for axions produced with larger initial energies. In the next section, we show that unless there exists a large hierarchy between the axion-photon and axion-nucleon coupling, $g_{a\gamma\gamma} / g_{aN} \ll 10^{-2}$, absorption is unlikely to quench the growth of the axion cloud. The axion-nucleon coupling is thus not expected to have a sizable impact in most axion models, and we therefore set $g_{aN} = 0$ in the rest of this paper.

\subsubsection{Self-interactions}
Non-linear corrections from the axion potential can lead to an alteration of the axion production rate. Taking the canonical instanton potential, $V(a) \propto (1-\cos(a/F_a))$, as a representative example, one can see that self-interactions become important if $a/F_a \sim \mathcal{O}(1)$. Here, $F_a$ is the axion decay constant. In order to remain in the linear regime, we are thus constrained to densities $\rho_{\rm L} \lesssim m_a^2 \, F_a^2$. Approximating $g_{a\gamma\gamma} \sim F_a^{-1}$, one can then estimate the typical density at which non-linearities become important. For an axion of mass $m_a = 10^{-6} \, {\rm eV}$ and an axion-photon coupling of $g_{a\gamma\gamma} = 10^{-11} \GeV^{-1}$, this density is around $\rho_{\rm NL} \sim 10^{33} \, {\rm GeV \, cm^{-3}}$. This is much larger than the typical densities reached in axion clouds (see next subsection); moreover, since the efficiency of axion production scales proportionally to $g_{a\gamma\gamma}^2$ and $\rho_{\rm NL}$ scales with $g_{a\gamma\gamma}^{-2}$, we do not expect non-linear densities to be reached at any mass or coupling. 

Self-interactions can further serve to quench the growth of the axion cloud via the $3 a \rightarrow a$ process. Here the final state axion has an energy $\omega_a \sim 3 \times m_a$, which is above the escape velocity of the neutron star. The rate of energy loss is given by~\cite{Arvanitaki:2010sy}
\begin{equation}
    \frac{dE_{3a\rightarrow a}}{dt} = 8\pi \omega_a^2 |\tilde{G}|^2 \, ,
\end{equation}
where
\begin{eqnarray}
    \tilde{G} = \frac{1}{24\pi}\frac{m_a^2}{F_a^2} \int d^3x \, e^{-i k \cdot x} \, a(x)^3 \, .
\end{eqnarray}
By equating the energy loss with the energy injection (here, we use the approximate expression for the energy injection derived in~\cite{PulsarNulling} in order to maintain the functional dependency on all quantities), we can determine the density and coupling at which axion emission could become important; this occurs at
\begin{multline}
    \rho_{3a\rightarrow a} \sim  7 \times 10^{26} \, \frac{{\rm GeV}}{{\rm cm^3}} \left(\frac{m_a}{10^{-6} \, {\rm eV}} \right)^{5/3} \\
    \times \left(\frac{5 \times 10^{-13} \, {\rm GeV^{-1}}}{g_{a\gamma\gamma}} \right)^{2/3} \left(\frac{B_0}{10^{12} \, {\rm G}} \right)^{4/3} \left(\frac{\Omega_{\rm NS}}{10 \, {\rm Hz}}\right)^{2/3} \, ,
\end{multline}
where for simplicity we have adopted characteristic values of the vacuum gap height and radius of $h_{\rm gap} \sim 10 \, \rm m$ and $r_{\rm pc} \sim 100 \, \rm m$ (see \eg~\cite{Noordhuis:2022ljw,PulsarNulling} for the full expressions of these quantities). As we show below, this density is never realizable, and thus $3a \rightarrow a$ axion emission cannot alter the growth of the axion cloud.

\subsubsection{Electromagnetic interactions}
Let us now turn our attention toward the impact of the electromagnetic coupling on the growth of the axion cloud. In general, this coupling can either $(i)$ cause axions to drive oscillatory electromagnetic fields, which dissipate energy into the local plasma, or $(ii)$ directly radiate low-energy photons. We discuss each of these possibilities below. 

The natural expectation is that energy stored in the axion cloud can only be efficiently dissipated prior to the discharge process; this is a consequence of the fact that the electric field induced by the axion, $\vec{E}_a$, is suppressed when the plasma frequency is larger than the axion mass~\cite{PulsarNulling}. In order to see this effect more explicitly, one can analyze the motion of an electron in the presence of an external electric field $\vec{E}_0$ and a background axion field. Following the derivation of~\cite{Beutter:2018xfx}, the transition amplitude of the electron (denoted by $\psi$, and carrying charge $-e$) between an initial state $\left. |i \right>$ and final state $\left. |f \right>$ at zeroth order (in axion-photon coupling) is given by
\begin{equation}
    \mathcal{A} = -i\left< f | \int d^4x \, \mathcal{H}_I(x) | i \right> = -i \int d^4x \, J^\mu(x) A_{\mu}^{\rm ext}(x) \, ,
\end{equation}
where the interaction Hamiltonian in the absence of the axion is given by $\mathcal{H}_I = - \mathcal{L}_{\rm em} =  e \bar{\psi} \gamma^\mu \psi A_{\mu}^{\rm ext}$, $J^\mu \equiv e \left< f|\bar{\psi}\gamma^\mu \psi|i\right>$, and $A_{\mu}^{\rm ext}$ is the vector potential due to an external electromagnetic field. Including the Lagrangian for the axion, \ie taking $\mathcal{H}_I \rightarrow -\mathcal{L}_{\rm em} - \frac{g_{a\gamma\gamma}}{4} \, a \, F \tilde{F}$, and working to first order, one finds
\begin{eqnarray}
\mathcal{A} &=& \frac{(-i)^2}{2!}  \left< f | \mathcal{T} \int d^4x \int d^4 y \, \mathcal{H}_I(x) \mathcal{H}_I(y) | i \right> \nonumber \\   
&=& - i \int d^4x \, J^\mu(x) A_\mu^{\rm ind}(x) \, .
\end{eqnarray}
Here $\mathcal{T}$ denotes time ordering, and the induced vector potential is given by
\begin{equation}
    A_\mu^{\rm ind}(x) = i g_{a\gamma\gamma} \int d^4y \, D_{\mu\nu}(x-y) \, \partial_\rho a(y) \, \tilde{F}^{\rho \nu}(y) \, .
\end{equation}
The standard dressed photon propagator (in the Feynman gauge) has the form
\begin{equation}
    D_{\mu\nu}(x-y) = \int \frac{d^4q}{(2\pi)^4} \, \frac{-i g_{\mu\nu}}{q^2 - \Pi_{L,T}} e^{-i q (x-y)} \, ,
\end{equation}
where $\Pi_{L,T}$ represents the longitudinal and transverse polarization tensors. In the non-relativistic limit, $\vec{E}_a \sim - \partial_t \vec{A}^{\rm ind}$, implying that in vacuum $\vec{E}_a \sim i g_{a\gamma\gamma} a \, \vec{B} \, e^{i (\vec{k}_a \cdot \vec{x} - m_a t)}$. In a dense plasma, where the axion mass and momentum are much smaller than the plasma frequency ($\omega_p$), one instead finds $\vec{E}_a \sim i g_{a\gamma\gamma} (m_a / \omega_p)^2 \, a \, \vec{B} \, e^{i (\vec{k}_a \cdot \vec{x} - m_a t)}$~\cite{PulsarNulling}. Here, the in-medium suppression of the electric field is apparent.

For sufficiently light axions (\ie axions that can be considered as roughly coherent during the acceleration phase of the gap), the phase of $\vec{E}_a$ is approximately constant over the gap. Depending on the sign of the phase, the axion-induced field will either boost or decelerate the primary charges. Without radiative losses, this effect would cancel, and there would be no energy transfer; instead, it is the emission of curvature radiation during the acceleration process which leads to non-zero energy loss of the axion cloud~\cite{PulsarNulling}. The rate of energy loss can be computed by noting that accelerating electrons radiate curvature photons at a rate given by~\cite{Muslimov2004}
\begin{equation}
    \dot{E}_c \sim \frac{2 e^2}{3 R_c^2}\gamma^4 \, ,
\end{equation}
where $R_c$ is the radius of curvature of the magnetic field lines and $\gamma$ is the boost factor of the electron. The total energy lost by all electrons to curvature radiation during the open phase can subsequently be approximated as
\begin{equation}
    \frac{dE_{\rm diss}}{dt} \sim A_{\rm gap} n_{\rm GJ} \int_0^{h_{\rm gap}} \, ds \, \dot{E}_c(s) \, , 
\end{equation}
with $A_{\rm gap} = \pi r_{\rm pc}^2$ being the area of the polar cap and $n_{\rm GJ}$ the GJ number density. The energy lost from the axion field that follows from this can be obtained by solving for the evolution of $\gamma(s)$ along a field line, and computing the relative energy loss in the presence of a perturbative correction to the bare electric field. This yields an average rate of energy loss given by~\cite{PulsarNulling}
\begin{equation}\label{eq:e_diss_low}
    \left< \frac{dE_{\rm diss}}{dt}\right> \sim \frac{g_{a\gamma\gamma}^2}{70\pi} \, \rho_a \, B_0^5 \, \Omega_{\rm NS}^3 \, \frac{e^5 \, h_{\rm gap}^7 \, r_{\rm pc}^2 \, \lambda_a^2}{m_e^4 \, R_c^2} \, \mathcal{F}^2 \, ,
\end{equation}
where $\rho_a$ is the axion energy density, $\lambda_a$ is the de Broglie wavelength of the axion, $m_e$ is the electron mass, and we have defined
\begin{eqnarray}
    \mathcal{F} \sim \begin{cases}
    1 \hspace{.05cm } & \lambda_a \ll h_{\rm gap}, r_{\rm pc}, R_{\rm NS} \\
    \left(\frac{h_{\rm gap}}{\lambda_a} \right) \hspace{.05cm } & h_{\rm gap} \ll \lambda_a \ll r_{\rm pc}, R_{\rm NS} \\
    \left(\frac{h_{\rm gap} r_{\rm pc}^2}{\lambda_a^3} \right) & h_{\rm gap}, r_{\rm pc} \ll \lambda_a  \ll R_{\rm NS} \\
    \end{cases} \, .
\end{eqnarray}
The factor $\mathcal{F}$ accounts for the relative in-medium suppression of the induced electric field that arises when the axion de Broglie wavelength is large relative to the gap (implying the induced electric field experiences a partial suppression due to the dense plasma outside the gap).

Eq.~\ref{eq:e_diss_low} applies only when the axion-induced electric field is coherent over the gap height. At larger axion masses, the incoherence of the induced field causes small oscillatory corrections to the evolutionary trajectory of the electron. The net potential difference, however, goes to zero in the large mass limit, and thus energy losses in this regime are heavily suppressed. Instead, for heavy axions, energy is most efficiently dissipated via the production of incoherent on-shell radiation (note that radiation is not produced at small axion masses since $e^\pm$ pair production, which kinematically blocks on-shell photon production, takes place on timescales below the oscillation length of the axion). The energy loss from photon production can be determined by computing the perturbed electric and magnetic fields, and looking at the Poynting flux in the far-field limit produced from these perturbations (see \eg Appendix D.c of~\cite{PulsarNulling} for a derivation using Green's functions) -- the rate of energy emission is given by~\cite{PulsarNulling}
\begin{equation}\label{eq:e_diss_high}
    \left< \frac{dE}{dt}\right>_{a\rightarrow \gamma} \simeq \frac{1}{4\pi} \frac{g_{a\gamma\gamma}^2 \, a^2 \, B_0^2}{\omega_a^2} \, \tau_{\rm duty} \, .
\end{equation}
Here $\tau_{\rm duty}$ is a duty cycle which has been included to account for the fact that photon emission occurs only when the gap is open. 

The axion energy density at saturation can be estimated by equating Eqs.~\ref{eq:e_diss_low} and~\ref{eq:e_diss_high} to the axion production rate (\ie the integral of Eq.~\ref{eqn:diffrate} over the momentum of local bound states). In order to derive analytic expressions for this, we use the approximations for the axion production rate computed in~\cite{PulsarNulling} (see \eg Appendixes C and D therein), which provide a rough estimate for the maximum achievable density of the axion cloud at the stellar surface. These saturation densities are given by~\cite{PulsarNulling}
\begin{equation}\label{eq:rho_sat1}
    \rho_{\rm sat} \sim 10^{24} \, \frac{\rm GeV}{\rm cm^3} \left(\frac{m_{a}}{10^{-6} \, \rm eV}\right)^3 \left[\frac{B_{0}}{10^{12} \, \rm G} \frac{\Omega_{\rm NS}}{{10 \, \rm Hz}}\right]^{9/7} \mathcal{F} \,
\end{equation}
for $\lambda_a \geq h_{\rm gap}$, and
\begin{multline}\label{eq:rho_sat2}
    \rho_{\rm sat} \sim {7 \times 10^{23}} \, \frac{\rm GeV}{\rm cm^3} \left(\frac{m_{a}}{10^{-6} \, \rm eV}\right)^3 \\ 
    \times \left(\frac{B_{0}}{10^{12} \, \rm G}\right)^{18/7} \left(\frac{\Omega_{\rm NS}}{10 \, \rm Hz}\right)^{39/7} \,
\end{multline}
for $\lambda_a < h_{\rm gap}$.

Eqs.~\ref{eq:rho_sat1} and~\ref{eq:rho_sat2} are derived under the assumption that the axion-induced electric field remains small relative to the electric field of the neutron star itself. If we look \eg at Gauss' law (in the corotating frame),
\begin{equation}
    \nabla \cdot \vec{E} = \rho_e - \rho_{\rm GJ} - g_{a\gamma\gamma} \, \vec{B} \cdot \nabla a \, ,
\end{equation}
one can see that this assumption breaks down when $\left< g_{a\gamma\gamma} \, \vec{B} \cdot \nabla a \right> \rightarrow \xi \rho_{\rm GJ}$. Here $\left< \cdot \right>$ denotes an averaging over the gap, and an order one `fudge factor' $\xi$ has been introduced in order to emphasize that the backreaction regime is not precisely known~\cite{PulsarNulling} -- in what follows we take $\xi = 0.1$ to ensure we conservatively avoid strong backreaction, although we emphasize that this may yield overly conservative results. The above comparison yields the axion energy density at which backreaction becomes important~\cite{PulsarNulling}
\begin{multline}\label{eq:br_rho}
    \rho_{\rm br} \sim 4 \times 10^{18} \, \frac{ {\rm GeV}}{ {\rm cm^3}} \left(\frac{\Omega_{\rm NS}}{10 \, \rm Hz}\right)^2 \left(\frac{5 \times 10^{-13} \, \rm GeV^{-1}}{g_{a\gamma\gamma}}\right)^2 \\
    \times \begin{cases}
        1 \hspace{.5cm} & {\lambda_a} \geq h_{\rm gap} \\
        (\lambda_a^{-1} \times h_{\rm gap})^2 \hspace{.5cm} & {\lambda_a} < h_{\rm gap}
    \end{cases} \, ,
\end{multline}
which can also be expressed as
\begin{multline}
    \rho_{\rm br} \sim 4 \times 10^{18} \, \frac{ {\rm GeV}}{ {\rm cm^3}}  \left(\frac{5 \times 10^{-13} \, \rm GeV^{-1}}{g_{a\gamma\gamma}}\right)^2 \\
    \times \begin{cases}
        \left(\frac{\Omega_{\rm NS}}{10 \, \rm Hz}\right)^2 \hspace{.5cm} & {\lambda_a} \geq h_{\rm gap} \\[10pt]
        \left(\frac{m_a}{1.5 \times 10^{-8} \, {\rm eV}} \right)^2 \, \left(\frac{\Omega_{\rm NS}}{10 \, \rm Hz}\right)^{6/7} \, \left( \frac{B_0}{10^{12} \, {\rm G}} \right)^{-8/7} \hspace{.5cm} & {\lambda_a} < h_{\rm gap} 
    \end{cases} \, .
\end{multline}
At higher densities the backreaction of the axion field becomes highly non-linear, and numerical simulations are likely required in order to understand whether the axion cloud can continue to grow. In order to remain conservative, in what follows we halt the growth of the cloud entirely when the axion density at the neutron star surface hits $\rho_{\rm max} \equiv {\rm Min}\left[\rho_{\rm br}, \, \rho_{\rm sat} \right]$\footnote{Given the large axion densities present in the gap, one may wonder whether axion production can be stimulated as \eg is the case in~\cite{Caputo:2018ljp,Caputo:2018vmy}. Using the formalism of~\cite{Caputo:2018vmy}, however, one can show that this does not occur; instead, large axion backgrounds eventually lead to absorption, driving the system to a state of equilibrium.}. Note that as the neutron star evolves we continue to enforce $\rho_a \leq \rho_{\rm max}(B_0, \Omega_{\rm NS})$. For the masses studied here, we find that $\rho_{\rm max}$ is almost always determined by the backreaction density.

A second prominent mechanism for electromagnetic energy dissipation in axion clouds is resonantly enhanced axion-photon mixing. This can occur when the local plasma frequency becomes comparable to the axion mass, \ie $\omega_p \simeq m_a$\footnote{This condition arises from determining the point at which the photon and axion four-momentum are equal, \ie $k_{\gamma}^\mu = k_a^\mu$. Here, the photon momentum corresponds to the Langmuir-O mode of a strongly magnetized plasma (see e.g.~\cite{Witte2021,McDonald:2023shx}). Note also that corrections from the Cotton-Mouton term~\cite{Hochmuth:2007hk,Fairbairn:2009zi} and the Euler-Heisenberg term~\cite{Raffelt:1987im,Dobrynina:2014qba} are entirely negligible for the systems of interest.}. Here, $\omega_p = \sqrt{4 \pi \alpha n_e / m_e}$, with $\alpha$ being the fine-structure constant and $n_e$ the electron/positron number density. Whether or not such a condition is actually met over the course of an axion trajectory depends crucially on both the properties of the magnetosphere and on the energy of the axion. Extremely low-energy axions are effectively confined to radii $r \lesssim R_{\rm NS} + h_{\rm gap}$, and for most neutron stars the plasma density is likely to be large enough to kinematically block the resonance in these regions\footnote{This is not necessarily true for the polar cap gaps themselves, as the plasma density there is naturally small in the vacuum phase. Nevertheless, this phase is expected to be too short to allow for meaningful resonant conversion, and thus likely will not be impactful in this regard.}. Axions with velocities near the escape velocity, however, will have large apoapsides, implying they always traverse from regions with $\omega_p \gg m_a$ to regions with $\omega_p \ll m_a$. Therefore, for these axions energy dissipation via resonant conversion is expected to be unavoidable. The net result is qualitatively different behavior for the growth of the axion cloud at small and large radii.

In order to provide an illustrative example of the impact of resonant axion-photon mixing on the evolution of the axion cloud, we adopt a spherically symmetric plasma distribution normalized to the GJ plasma frequency at the pole (\ie the $e^\pm$ number density at the pole is set to $n_e \equiv 2 \vec{B}_0 \cdot \vec{\Omega}_{\rm NS} / e$)~\cite{GoldreichJulian1969}, and with radial dependence scaling like $(r/R_{\rm NS})^{-3/2}$. Using this charge distribution, one can determine that resonant axion-photon transitions occur at a fixed radius $r_c$ given by
\begin{equation}\label{eq:conv_rad}
    r_c = 25 \, \textrm{km} \left(\frac{R_{\rm NS}}{10 \, \rm km}\right) \left[\frac{B_0}{10^{12} \, \rm G} \frac{\Omega_{\rm NS}}{1 \, \rm Hz} \left(\frac{10^{-6} \eV}{m_a}\right)^2 \right]^{1/3} \, .
\end{equation}
The efficiency of conversions is expected to be roughly~\cite{Safdi:2018oeu,Battye:2019aco,Witte2021,millar2021axionphotonUPDATED,mcdonald2023axionphoton}
\begin{equation}\label{eq:conv_prob}
    P_{a \rightarrow \gamma} \simeq 1 - \textrm{exp}\left[-\frac{\pi}{2} \frac{g_{a\gamma\gamma}^2 B_c^2}{v_c \left| \omega_p' \right|}\right] \, ,
\end{equation}
where $v_c$ and $B_c$ are the axion velocity and magnetic field at the conversion radius, and $\omega_p'$ is the gradient of the plasma frequency projected onto $\hat{v}_c$. For simplicity, we have assumed in Eq.~\ref{eq:conv_prob} that the magnetic field is always perpendicular to the axion velocity at conversion. The angular factors correcting for non-perpendicular propagation are not necessary for the rough estimates performed here, since the photon flux quickly becomes independent of conversion probability (this occurs due to the growth of the axion cloud reaching a state of equilibrium -- see following paragraphs for more details). 

\begin{figure}
    \centering
    \includegraphics[width=0.48\textwidth]{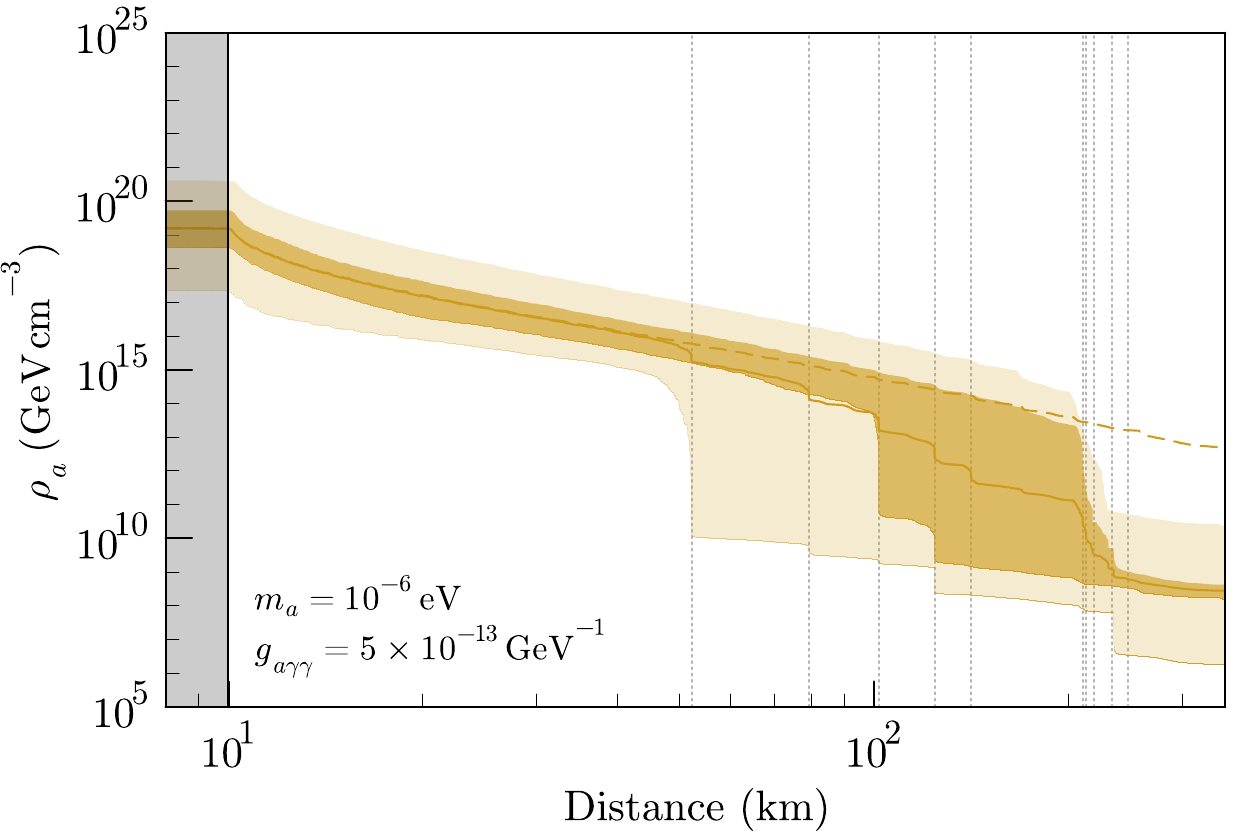}
    \caption{Axion cloud density profile distribution after $1 \, \rm kyr$, and including continuous resonant transitions from a spherical conversion surface as defined in Eq.~\ref{eq:conv_rad}. The distribution mean is plotted along with its interquartile range (dark shaded area), minimum, and maximum (light shaded area). The constant neutron star radius is shown as a solid vertical line, with the area inside of the neutron star being grayed out. The density profile mean without resonant conversions is furthermore included as a dashed line, and the conversion radii of all considered neutron stars are shown as dotted vertical lines.}
    \label{fig:Edensity}
\end{figure}

\begin{figure*}
    \centering
    \includegraphics[width=0.96\textwidth]{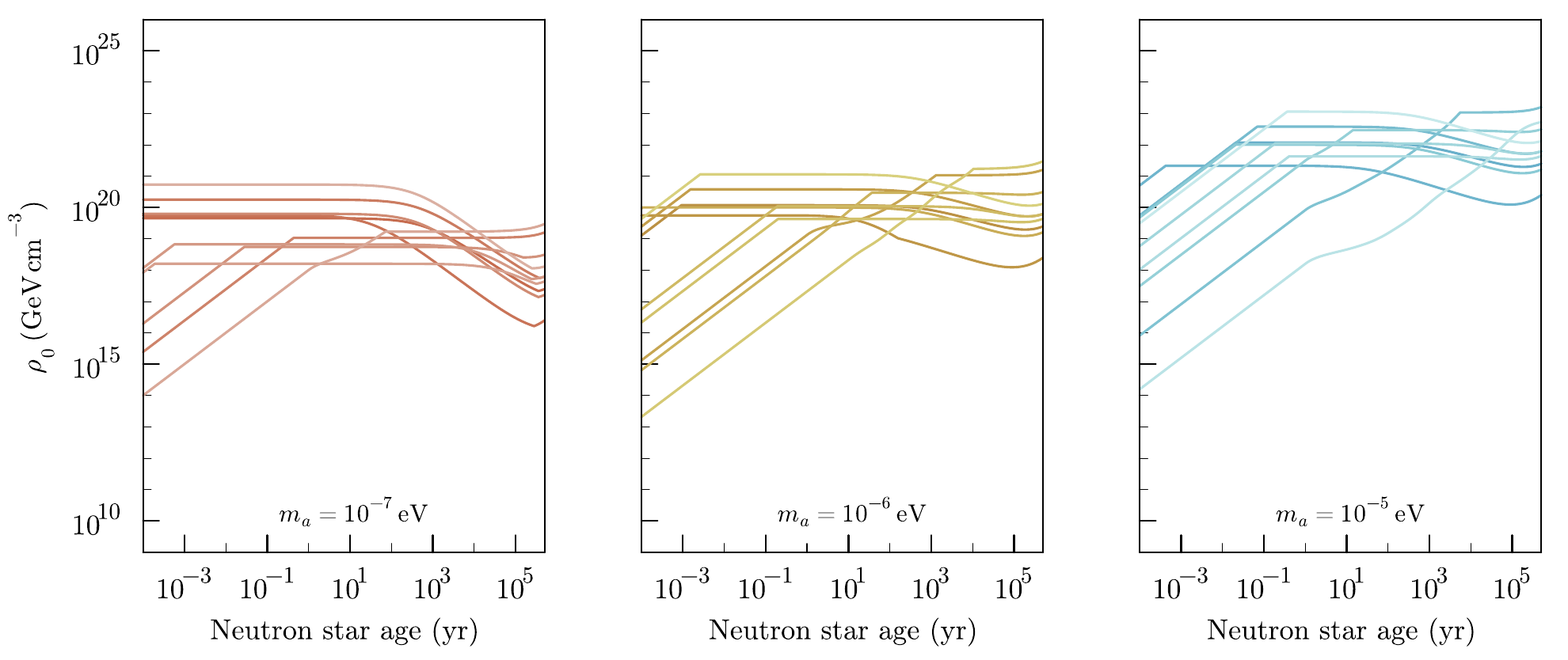}

    \caption{Axion cloud density at the neutron star surface as a function of neutron star age, plotted for our ten neutron star samples at three different values of the axion mass. The axion-photon coupling is set to $g_{a\gamma\gamma} = 5 \times 10^{-13} \GeV^{-1}$. }
    \label{fig:EdenEvol}
\end{figure*}

\begin{figure*}
    \centering
    \includegraphics[width=0.96\textwidth]{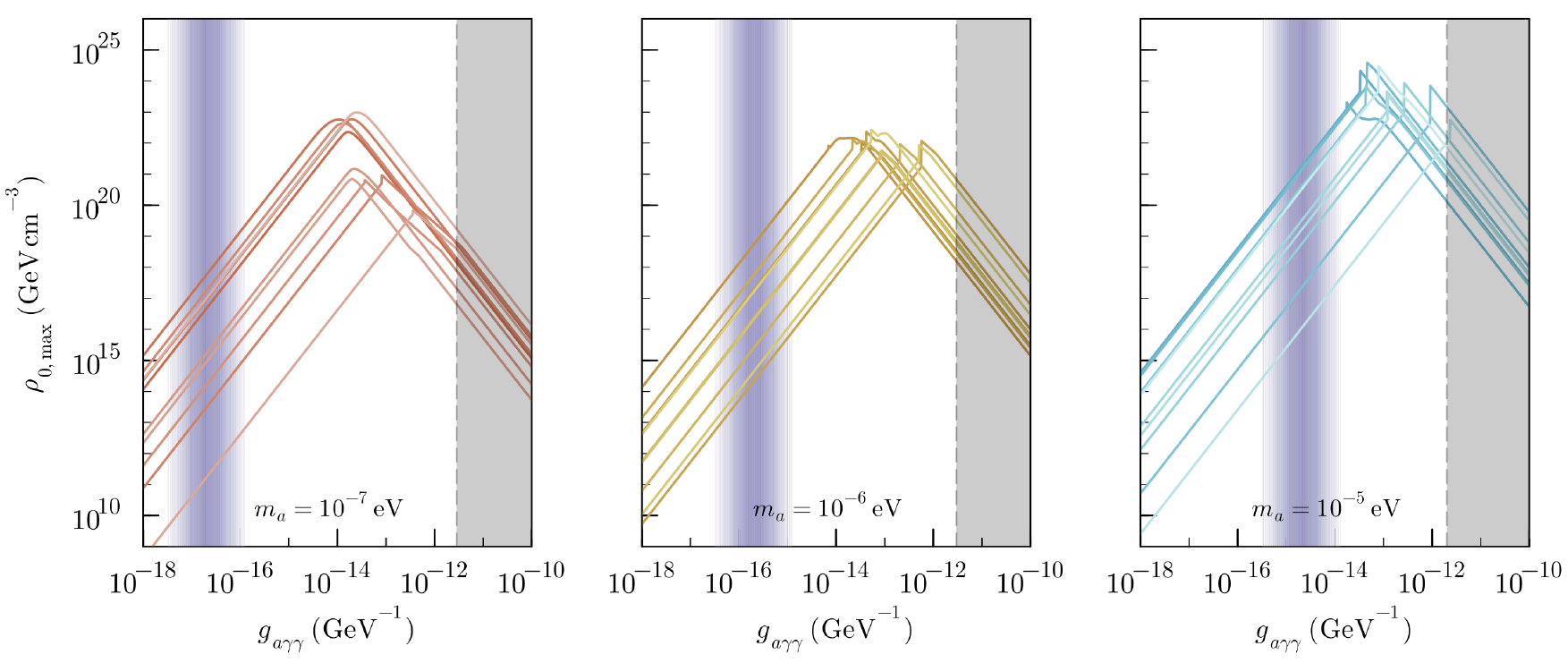}
    \caption{Maximal axion density achieved at the neutron star surface as a function of axion-photon coupling, plotted for our ten neutron star samples at three different values of the axion mass. The approximate parameter space associated to the QCD axion is shown using vertical purple bands~\cite{DiLuzio:2020wdo}. Previously excluded regions of parameter space~\cite{Noordhuis:2022ljw} are shaded in gray. We note that the behavior at high couplings is a result of truncating the density at $\rho_{\rm max}$, which has been implemented in order to conservatively avoid evolving the axion cloud in the regime where the axion can backreact on the electrodynamics. }
    \label{fig:Pole_Edensity}
\end{figure*}

Note that, unlike previous studies in this field, we have chosen not to use the charge-separated GJ model; this model contains regions of pure vacuum extending to the surface of the neutron star (occurring when $\vec{B}_0 \cdot \vec{\Omega}_{\rm NS} = 0$), a feature which is likely not physical for young pulsars and which would dramatically enhance radio emission (the choice to remove these features thus leads to conservative predictions of the radio flux\footnote{We have verified using the neutron star parameters at death that the choice of a spherical conversion surface can suppress the radio flux by as much as 4 orders of magnitude (although this suppression is significantly reduced for large axion masses, when the conversion surface is generally close to the neutron star surface either way).}). Note, however, that full charge separation is expected to occur in dead neutron stars (see \eg~\cite{Safdi:2018oeu}) -- we return to the implications of this transition in the following section. 
 
Because of the spherical conversion surface, low-energy axions, which are confined to orbits with apoapsis $r \leq r_c$, will never dissipate their energy. Higher-energy axions, on the other hand, will cross the conversion surface many times, converting with a low probability at each crossing. In order to assess the effect of these resonances on the axion cloud, we identify how frequently resonances are encountered and how efficient each resonance is at depleting the axion cloud. Conversions are anticipated to reach an equilibrium, where the rate of axion injection matches the rate of energy dissipation -- this equilibrium roughly occurs on a timescale $\tau_{\rm eq} \sim \tau_{\delta{\rm res}} / \left <P_{a\rightarrow \gamma}\right>$, where $\tau_{\delta{\rm res}}$ is the timescale between resonant crossings and $\left <P_{a\rightarrow \gamma}\right>$ is the typical fraction of axions lost at each crossing.

Fig.~\ref{fig:Edensity} shows the reconstructed density profiles of our ten neutron star samples after $1 \, \rm kyr$, adopting fiducial axion parameters $m_a = 10^{-6} \eV$ and $g_{a\gamma\gamma} = 5 \times 10^{-13} \GeV^{-1}$, and taking the neutron star radius and mass to be $R_{\rm NS} = 10 \km$ and $M_{\rm NS} = M_\odot$, respectively. Note that unless otherwise indicated, these values are used as fiducial parameters throughout the text. The dark and light bands reflect the variation across the neutron star population, with the darker bands showing the $\pm 25\%$ quartiles about the mean, and the lighter bands the minimum/maximum of the population. The density profile exhibits notable dips at radii near $\mathcal{O}(100)\, \rm km$, a result of the fact that energy in the bound states here is being dissipated via resonant photon production. The density profile at smaller radii is unaffected by resonant conversions, and will therefore grow unimpeded. In the absence of conversions, the reconstructed slope outside of the neutron star is found to scale like $\rho \propto r^{-4}$ \footnote{Inside the neutron star we instead find that the density is approximately constant. We however caution the reader that in order to extend trajectories inside of the neutron star we have simply applied a rescaling to the neutron star mass which assumes a constant internal density. In general, one should use the internal Schwarzschild metric in this regime, and thus our scaling may not be robust.}.

In order to highlight the importance of $\rho_{\rm max}$, we plot in Fig.~\ref{fig:EdenEvol} the time evolution of the axion energy density at the neutron star surface for three axion masses. Here, each line corresponds to a single neutron star in our sample. Fig.~\ref{fig:EdenEvol} clearly shows sharp interruptions in linear growth of the axion density, reflecting the time at which the axion cloud achieves a surface density equal to $\rho_{\rm max}$ (we remind the reader that the maximal density is generally set by the backreaction density given in Eq.~\ref{eq:br_rho}). Note that this saturation can occur across a range of different timescales spanning from $\tau \ll 10^{-4} \, \rm yrs$ to $10^6 \, \rm yrs$, depending on the axion parameters and neutron star. The non-linear evolution seen in these curves at late times reflects the impact of magnetorotational spin-down, which can affect both the value of the backreaction density as well as the axion production rate.

In Fig.~\ref{fig:Pole_Edensity} we illustrate the scaling of the axion density with axion-photon coupling, plotting the maximal surface density achieved over the course of the neutron star lifetime as a function of $g_{a\gamma\gamma}$. For sufficiently large couplings, the backreaction density in Eq.~\ref{eq:br_rho} causes the growth to saturate early in a neutron star's life; since Eq.~\ref{eq:br_rho} scales with  $g_{a\gamma\gamma}^{-2}$, the maximal density in the axion cloud grows at smaller couplings. There is a turnover (\ie the sharp peaks seen in Fig.~\ref{fig:Pole_Edensity}), however, which for typical neutron stars occurs between $10^{-14} \, {\rm GeV^{-1}} \lesssim g_{a\gamma\gamma} \lesssim 10^{-12} \, {\rm GeV^{-1}}$ (also depending on the axion mass) -- below this threshold, $\rho_{\rm max}$ is not reached within the neutron star's lifetime, and thus the maximally achieved density in this regime simply scales with $g_{a\gamma\gamma}^{2}$. Fig.~\ref{fig:Pole_Edensity} highlights two important points. Firstly, there is a general trend that larger axion masses achieve higher densities, however, the peak density also tends to occur at larger couplings. Secondly, large axion densities, which exceed the local dark matter density by more than 10 orders of magnitude, are achieved at all couplings for the mass range studied here, including \eg QCD axion parameter space (which has been highlighted using the purple band in Fig.~\ref{fig:Pole_Edensity}).

\section{Radio Emission}\label{sec:radio}
Sec.~\ref{sec:formation} establishes that enormous axion energy densities can be achieved around neutron stars via the gap-production process. The dissipation of this energy via resonant photon production provides a direct observable to probe the existence of axions, namely radio emission. Here we investigate signatures arising from this process, and in Sec.~\ref{sec:sensitivity} we discuss the extent to which they are observable using existing radio infrastructure. 

\begin{figure}
    \centering
    \includegraphics[width=0.48\textwidth]{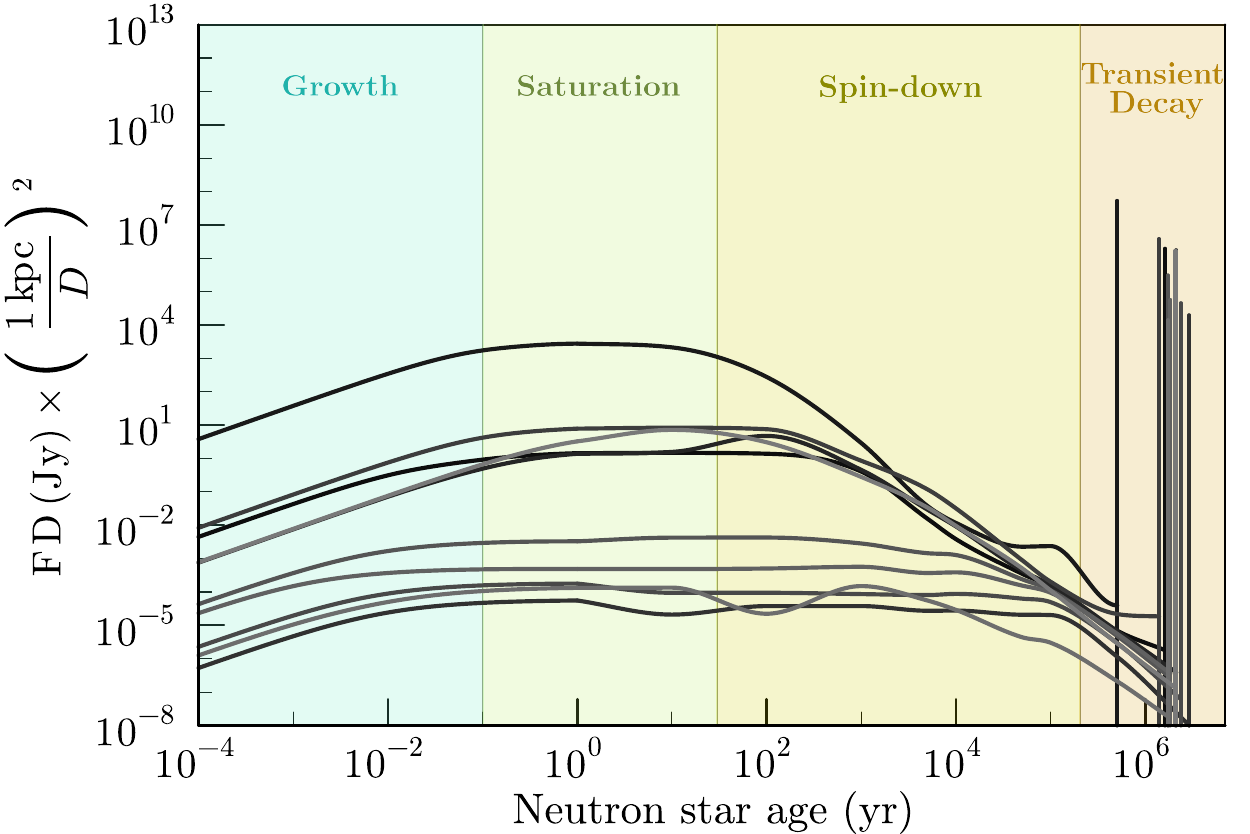}
    \caption{Temporal evolution of the flux density (FD) (produced via resonant axion-photon mixing) from the ten neutron stars in our sample, assuming a distance of $1 \, \rm kpc$. Four emission phases are identified: linear growth, saturation, spin-down, and transient decay (see text for details on the physics driving each evolutionary stage). Results are shown for fiducial values of the axion mass and axion-photon coupling, $m_a = 10^{-6} \, {\rm eV}$ and $g_{a\gamma\gamma} = 5 \times 10^{-13} \, {\rm GeV}$. }
    \label{fig:FluxEvol}
\end{figure}

In order to provide a rough understanding of the magnitude and time evolution of the radio flux,
we follow the change in the axion production rate and the energy dissipation rate over the lifetime of each of the neutron stars in our sample. Assuming a characteristic distance of $1 \, \rm kpc$ and fixing the bandwidth to be the minimal frequency range capturing $95 \%$ of the flux (which typically corresponds to $1 - 10 \MHz$), we show in Fig.~\ref{fig:FluxEvol} the evolution of the flux density for our fiducial parameters. Here we see that, like the axion cloud itself, the radio emission of each neutron star undergoes four distinct phases of evolution: growth, saturation, spin-down, and transient decay. We outline the physics driving each of these phases below. 

{\bf Growth:} After the neutron star settles into its relaxed state, pair production processes in the polar cap begin producing a non-relativistic population of axions. So long as the axion density remains sufficiently low so that energy dissipation is small, the axion cloud and the radio flux grow linearly with time. In this regime, the radio flux scales proportional to $g_{a\gamma\gamma}^4$, with two factors coming from axion production and two coming from resonant conversion. The radio flux traces the evolution of the axion density until the profile equilibrates, a process which occurs at the point when bound state axions are converted with the same intensity as they are injected.

\begin{figure*}
    \centering
    \includegraphics[width=0.90\textwidth]{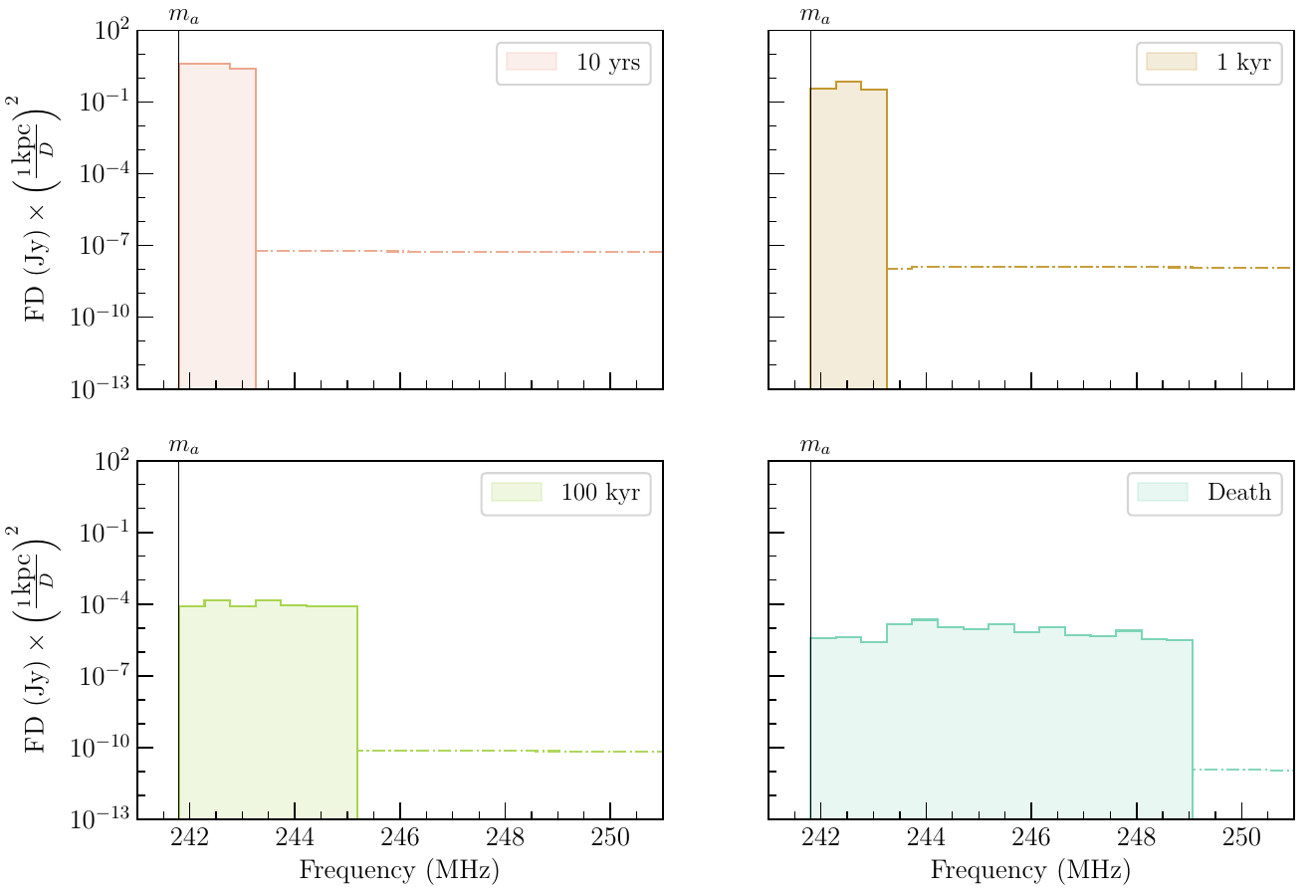}

    \caption{Spectrum of radio emission generated from axion bound states (solid lines, filled areas) and the non-bound axion population (dot-dashed lines). Results are shown for neutron star 5 (see Table~\ref{tab:pulsars}) at four different ages. The axion mass and axion-photon coupling are respectively set to $m_a = 10^{-6} \eV$ and $g_{a\gamma\gamma} = 5 \times 10^{-13} \GeV^{-1}$. }
    \label{fig:Spectra2}
\end{figure*}

{\bf Saturation:} Once equilibrium is reached, the radio flux and spectrum will remain approximately constant until the neutron star begins to spin down. Note that the inner part of the density profile does continue to grow, as low-energy axions are not expected to cross resonant conversion surfaces. During this saturation phase, the axion-photon conversion probability becomes {\emph{independent}} of $g_{a\gamma\gamma}$ and therefore the radio flux scales as $g_{a\gamma\gamma}^2$ rather than $g_{a\gamma\gamma}^4$. This implies that radio emission from bound states is amplified relative to the radio emission generated from axions with $k_a > k_{\rm esc}$ (\ie the axion population studied in~\cite{Noordhuis:2022ljw}) by a factor of $1/\left< P_{a\rightarrow\gamma}\right>$. Note that the photon energy is inherited directly from the axion distribution at the resonant conversion surface, and as such one expects the radio spectrum to be sharply enhanced at frequencies $m_a \lesssim \omega \lesssim \sqrt{k_{\rm esc}(r_c)^2 + m_a^2}$. A secondary distinctive feature in the spectrum arises from the kinematic threshold set by the axion mass -- this provides a sharp lower limit for the peaked radio emission. Assuming momentarily that the conversion takes place at characteristic radii of $r_c \sim 10-100 \times R_{\rm NS}$, one thus expects all photons to be produced with energies located within a few percent of the axion mass (an absolute upper limit on the energy width at production, coming from assuming a $\sim 2.2 \, M_\odot$ neutron star with conversion occurring directly at the surface, is around $\sim 20\%$). It is worth mentioning that gravitational redshifting during a photon's escape will induce an additional $\sim 1\% - 10\%$ level shift in in the location of the line\footnote{Note that the variation in the gravitational potential over the gap itself typically contributes only at the level of $\mathcal{O}(0.1 \%)$, and is thus highly subdominant.}. For simplicity, we neglect this energy shift in our calculations; however, it is a trivial effect to incorporate in future analyses. Collectively, the above characteristics serve to produce an overall sharp end-point in the radio spectrum.

We illustrate this in Fig.~\ref{fig:Spectra2} using one of the sampled neutron stars at various ages. One can see that the radio flux generated by bound states (shaded region) is clearly elevated with respect to the flux produced by axions that escape the magnetosphere (dot-dashed lines). We do note that the instantaneous nature of the transition at the high-frequency edge is a consequence of adopting a perfectly spherical conversion surface, but even in more realistic scenarios the drop is expected to remain sharp. Fig.~\ref{fig:Spectra2} furthermore shows that both the width and amplitude of the spectral feature evolve over the course of the neutron star's life -- this is a direct consequence of the fact that magnetorotational spin-down shifts the conversion surface to smaller radii while simultaneously decreasing the injection rate of axions, modifying both the frequency and characteristic conversion probability of produced photons.

\begin{figure*}
    \centering
    \includegraphics[width=0.96\textwidth]{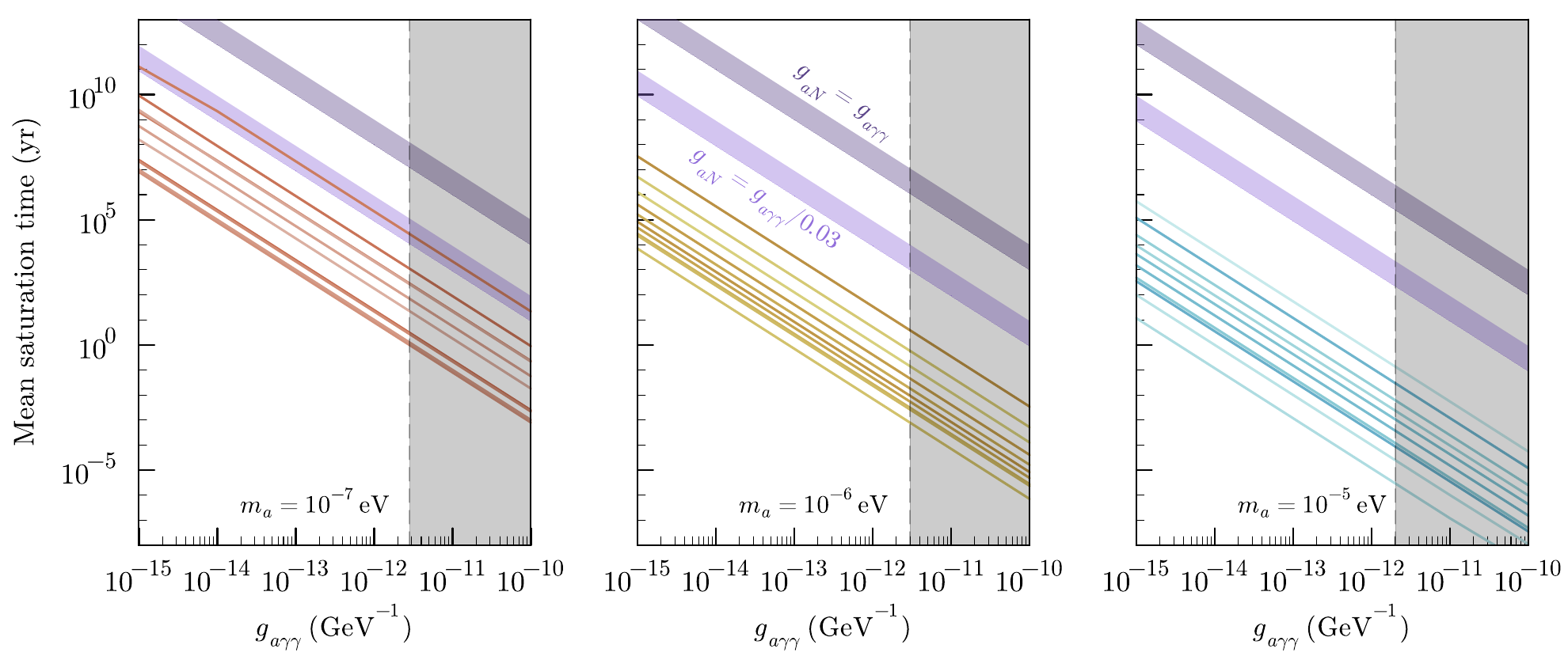}

    \caption{Mean saturation time (\ie the average time it takes before the axions that contribute to the resonant flux have a conversion rate equal to $90 \%$ of their production rate) as a function of axion-photon coupling, plotted for our ten neutron star samples at three different values of the axion mass. Previously excluded regions of parameter space~\cite{Noordhuis:2022ljw} are shaded in gray. For comparison, we also show for each neutron star the mean absorption timescale (\ie the average time it takes before $90 \%$ of produced axions are being absorbed inside of the neutron star) for axions whose apoapsis is fixed to the conversion radius -- to avoid clutter, this is displayed as a band, rather than ten individual lines. Absorption timescales are shown for two characteristic ratios of $g_{aN} / g_{a\gamma\gamma}$, one corresponding to equality and the other to the ratio for the QCD axion in the KSVZ model (see \eg~\cite{Irastorza:2018dyq}). }
    \label{fig:SatTime}
\end{figure*}

The characteristic time required in order to reach saturation depends on both the axion mass and the axion-photon coupling; we provide an estimation of this saturation time in Fig.~\ref{fig:SatTime}. This figure shows that saturation is not guaranteed to occur for sufficiently small couplings, as neutron stars may not always reach equilibrium before crossing the death line. We also plot in Fig.~\ref{fig:SatTime} the typical axion-nucleon bremsstrahlung absorption timescale, $\tau_{90}$, adopting an axion-nucleon coupling of $g_{aN} = g_{a\gamma\gamma}$ (dark purple) and $g_{aN} = g_{a\gamma\gamma} / 0.03$ (light purple), the latter roughly corresponding to the expected ratio in the KSVZ model -- these curves illustrate that axion absorption is generally insubstantial compared to resonant axion-photon mixing, and thus unlikely to be important in the evolution of the axion cloud.

{\bf Spin-down:} Magnetorotational decay of the neutron star causes the rotational period to increase and the magnetic field to decrease. This leads to two counteracting effects:
\begin{itemize}
    \item The characteristic charge density in the magnetosphere tends to decrease, implying a shift in the resonant conversion radius to smaller values where axion densities are larger. This serves to increase the radio flux.
    \item The production rate of axions tends to decrease, which translates to a decrease in the radio flux.
\end{itemize}
We have found that the latter effect generally controls the long-term evolution of the flux density, resulting in a diminishing flux as a function of neutron star age during this phase. Nevertheless, small deviations from this trend can occur, as seen in Fig.~\ref{fig:FluxEvol}.

{\bf Transient Decay:} The magnetospheres of dead neutron stars are known to relax to the so-called electrosphere solution (also referred to as the disk-dome solution), which is a fully charge-separated state that resembles the GJ model near the surface of the neutron star but falls off more quickly at large radii (see \eg~\cite{Safdi:2018oeu}).

Within this solution, regions of vacuum open in between the regions of positive and negative charges (extending all the way to the neutron star surface), allowing for axions of all energies (\ie at all radii) to resonantly convert to radio photons. Since pair production processes are no longer efficient in the polar caps of dead neutron stars, axion production is furthermore expected to cease. This implies that all residual energy stored in the axion cloud will dissipate -- in other words, dead neutron stars {\emph{typically}} do not have axion clouds\footnote{For axion masses $m_a \geq 10^{-5} \eV$ we find that in some cases the conversion surface shrinks below the neutron star radius before the magnetosphere charge separates; if this happens there will be no transient phase, and the neutron star will be left with a residual axion cloud that slowly dissipates via non-resonant mixing (for which the decay timescale can be quite sizable, see \eg~\cite{PulsarNulling}).}. Assuming that vacuum regions near the neutron star surface open on the light-crossing time, the timescale over which energy dissipates can be rapid, potentially generating a transient burst or perhaps a slower transient decay. The exact properties of the transient phase depend heavily on the axion mass, axion-photon coupling, and the strength of the neutron star's magnetic field.

\begin{figure*}
    \centering
    \includegraphics[width=0.96\textwidth]{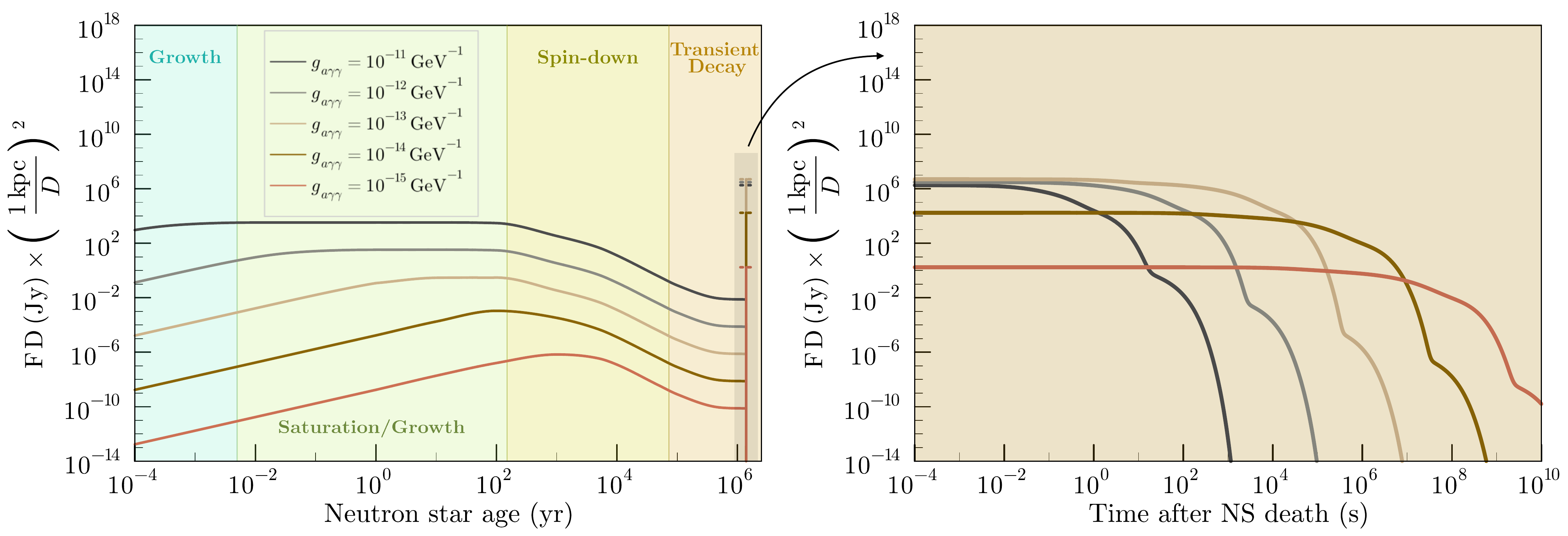}

    \caption{{\it Left:} Temporal evolution of the flux density (produced via resonant axion-photon mixing) from neutron star 5 (see Table~\ref{tab:pulsars}), plotted as a function of axion-photon coupling. The emission phases are the same as in Fig.~\ref{fig:FluxEvol}. {\it Right:} Zoom-in on the transient burst. The axion mass is set to $m_a = 10^{-6} \eV$. }
    \label{fig:FluxEvol_1NS}
\end{figure*}

\begin{figure}
    \includegraphics[width=0.48\textwidth]{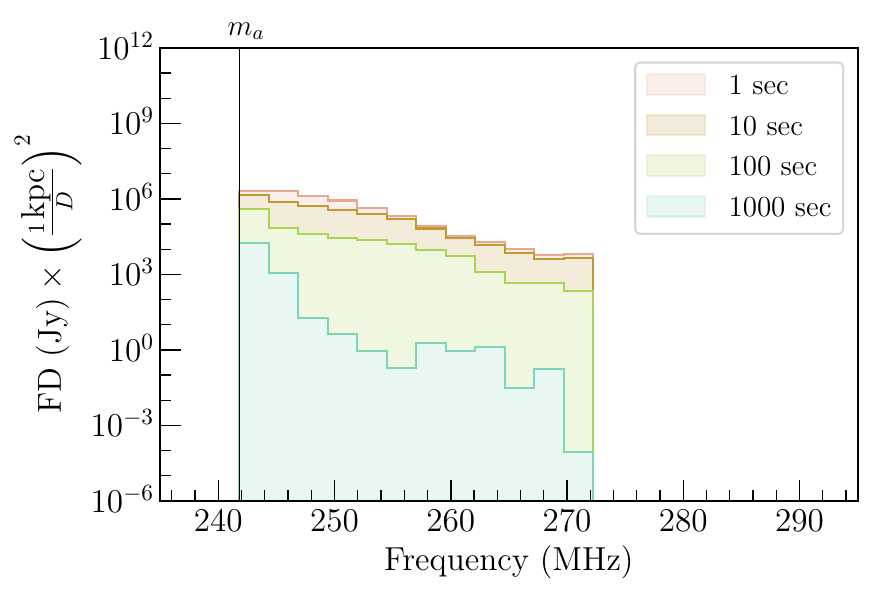}
    \caption{Transient spectrum of neutron star 5 (see Table~\ref{tab:pulsars}) at four different times after charge separation (assumed to occur instantaneously when the neutron star crosses the death line). The axion mass and axion-photon coupling are respectively set to $m_a = 10^{-6} \eV$ and $g_{a\gamma\gamma} = 5 \times 10^{-13} \GeV^{-1}$. }
    \label{fig:Spectra}
\end{figure}

\begin{figure*}
    \centering
    \includegraphics[width=0.96\textwidth]{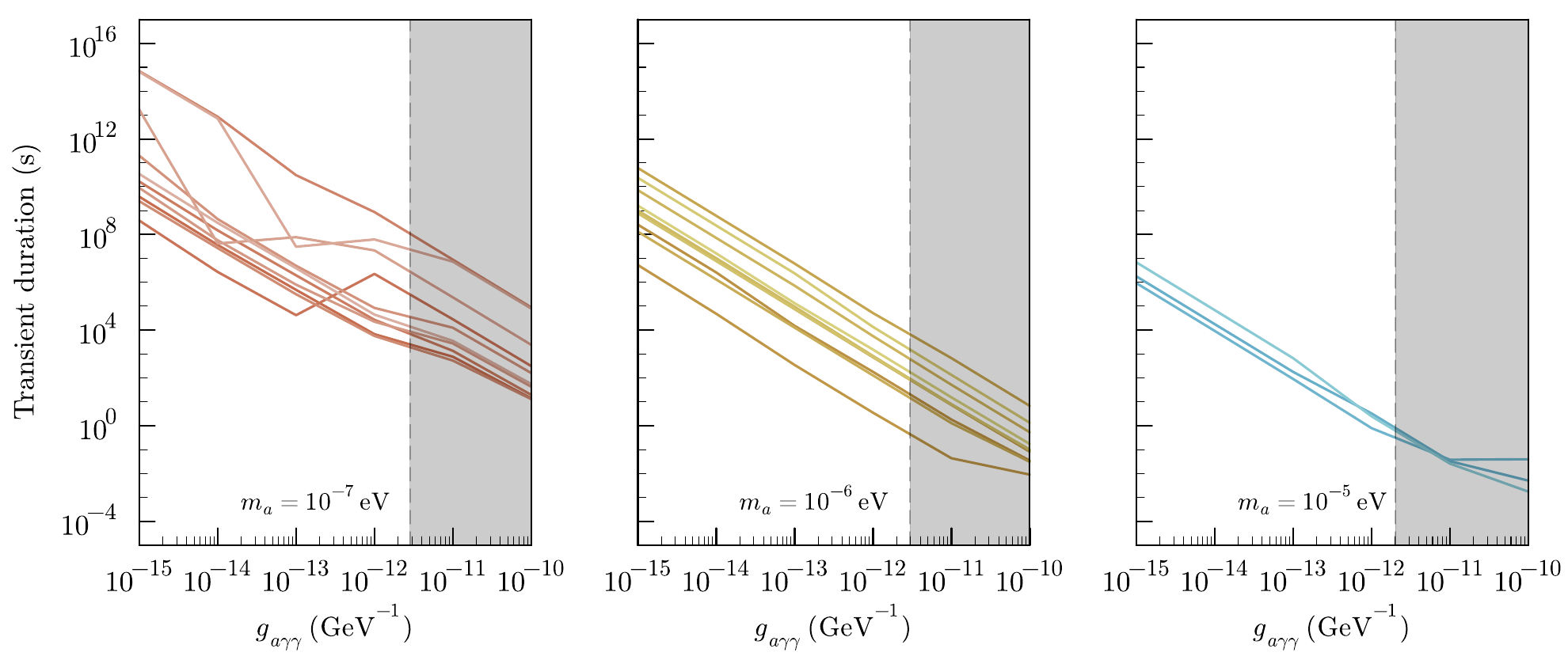}
    
    \caption{Transient duration as a function of axion-photon coupling, plotted for our ten neutron star samples at three different values of the axion mass. At $m_a = 10^{-5} \eV$ only three neutron stars are shown because the other samples do not produce a transient (as a result of their conversion surface shrinking below the neutron star radius before magnetosphere charge separation). Previously excluded regions of parameter space~\cite{Noordhuis:2022ljw} are shaded in gray. }
    \label{fig:TransTime}
\end{figure*}

Here, we model the transient by assuming the plasma density near the neutron star relaxes approximately instantaneously (\ie within the light-crossing time of the magnetosphere) to the charge-separated GJ charge density, given by
\begin{equation}
n_{\rm GJ} \simeq \frac{2 \vec{B}_0 \cdot \vec{\Omega}_{\rm NS}}{e} \, .
\end{equation}
Using the traced trajectories, and assuming the axion production mechanism has been abruptly turned off, we subsequently determine the rate at which energy is dissipated from the cloud. For large masses and couplings, this process may dissipate the entire axion cloud over timescales on the order of a second, while at lower masses and couplings the transient phase may last more than thousands of years. 

The temporal evolution of the transient as a function of coupling is illustrated for a particular neutron star in Fig.~\ref{fig:FluxEvol_1NS}. The zoom-in here clearly shows a break in the flux density profile, which arises because there exist two axion subpopulations within the cloud (differentiated by the sharp dip in the density profile, see Fig.~\ref{fig:Edensity}). Those axions that were previously continuously converting dissipate their energy on a relatively long timescale, and generally contribute only a small part of the initial transient flux (this is a result of the fact that the growth of this part of the axion cloud quenches at lower densities due to radiative emission). Conversely, the axions closer to the neutron star dissipate on shorter timescales, and tend to generate a larger flux density. In general, one expects this latter flux density to scale as $g_{a\gamma\gamma}^4$ (since larger couplings imply both a larger axion density and faster energy dissipation). However, for sufficiently large couplings, there are two effects that cause a deviation from this expectation. Firstly, and most prominently, we find that the buildup of the axion cloud eventually becomes limited by the backreaction density reported in Eq.~\ref{eq:br_rho}. This results in the achieved cloud density starting to scale with $g_{a\gamma\gamma}^{-2}$ instead of $g_{a\gamma\gamma}^2$. Secondly, one enters the adiabatic regime in which the conversion probability can saturate to $P_{a\rightarrow \gamma} \sim 1$, further softening the functional dependence on the coupling. Together, these effects can cause the transient flux density to slightly decrease at larger axion-photon couplings. Finally, we note that at even higher couplings (couplings which, albeit, are typically excluded by other constraints) it is occasionally possible for the high-energy axion population to dominate the flux, since the flux produced by this population always scales as $g_{a\gamma\gamma}^2$. For completeness, we also plot the evolution of the transient spectrum in Fig.~\ref{fig:Spectra} using the same parameters and neutron star as in Fig.~\ref{fig:FluxEvol_1NS}. Here, one can see that the transient burst occurs over a narrow window of frequencies and contains a slight spectral tilt, peaking near the axion mass.

The transient duration, defined here to be the time over which $90\%$ of the energy is dissipated, is shown for each of our neutron stars in Fig.~\ref{fig:TransTime} as a function of axion-photon coupling. Notice first of all that at $m_a = 10^{-5} \eV$ most neutron stars do not experience the transient event -- this is because the characteristic plasma density at death is well below the axion mass. Fig.~\ref{fig:TransTime} also shows the presence of multiple spectral breaks, causing the curves to deviate from a simple power-law scaling. At large masses this occurs when the conversion becomes adiabatic (\ie $P_{a \rightarrow \gamma}$ approaches order one values), while at small masses a dip feature appears due to the interplay between the two axion subpopulations. Here, the transient flux density generated by both populations is relatively comparable, and as a result the transient duration, which is set by the relative dissipation timescales of the two populations, can scale in a slightly non-linear way with the coupling. This materializes as a dip in Fig.~\ref{fig:TransTime}, which clearly manifests in a subset of the neutron stars in the population. The other stars show a drop in transient duration only when reaching lower couplings, with the subsequent rise expected to occur outside of the plotting region.

Having described the general spectral features in the radio band that are expected to arise from the evolution of axion clouds, we now turn our attention to the extent to which these features are observable. We caution the reader that these projections should be interpreted with care; despite the fact that we expect the qualitative evolution of these systems to be relatively robust, the choices made to describe the pulsar population and the evolution of the charge distribution can alter the quantitative estimates performed here. We reserve a more thorough investigation into the impact of these systematics on the projected sensitivity for future work.

\section{Sensitivity}\label{sec:sensitivity}
The radio flux from a pulsar observed on Earth is given by $F = dP/d\Omega/D^2$, where $dP/d\Omega$ is the differential power emitted from the pulsar along the line-of-sight and $D$ is the distance to the pulsar. Sensitivity of radio telescopes is often quoted in terms of flux density, defined as $S=F/\Delta \nu$, where $\Delta \nu$ is the signal bandwidth. The minimum detectable flux density for a given signal-to-noise ratio (SNR) of a single radio telescope is given by (see \eg~\cite{Cordes2003})
\begin{align}\label{eqn:telescope_sensitivity}
    S_{\rm min} = {\rm SNR} \times \frac{T_{\rm syst}}{G \sqrt{N_{\rm pol} \tau \Delta \nu}} \, .
\end{align}
Here $T_{\rm syst}$ is the system temperature, which accounts for all sources of astrophysical backgrounds and receiver noise. The telescope `gain', $G$, describes the effective collecting area of the telescope, $N_{\rm pol}$ is the number of polarization states being observed, and $\tau$ is the signal integration time. Radio telescope arrays combine observations from many telescopes by correlating the observations along the independent baselines established by $N_{\rm tele}$ independent instruments; collectively, this procedure allows for an enhancement in sensitivity set by (see \eg~\cite{Caputo:2018vmy})
\begin{equation}\label{eq:array}
S_{\rm min}^{\rm array} = \frac{S_{\rm min} }{\sqrt{\frac{1}{2} N_{\rm tele} (N_{\rm tele} - 1) }} \, .
\end{equation}

\begin{figure*}
    \centering
    \includegraphics[width=0.49\textwidth]{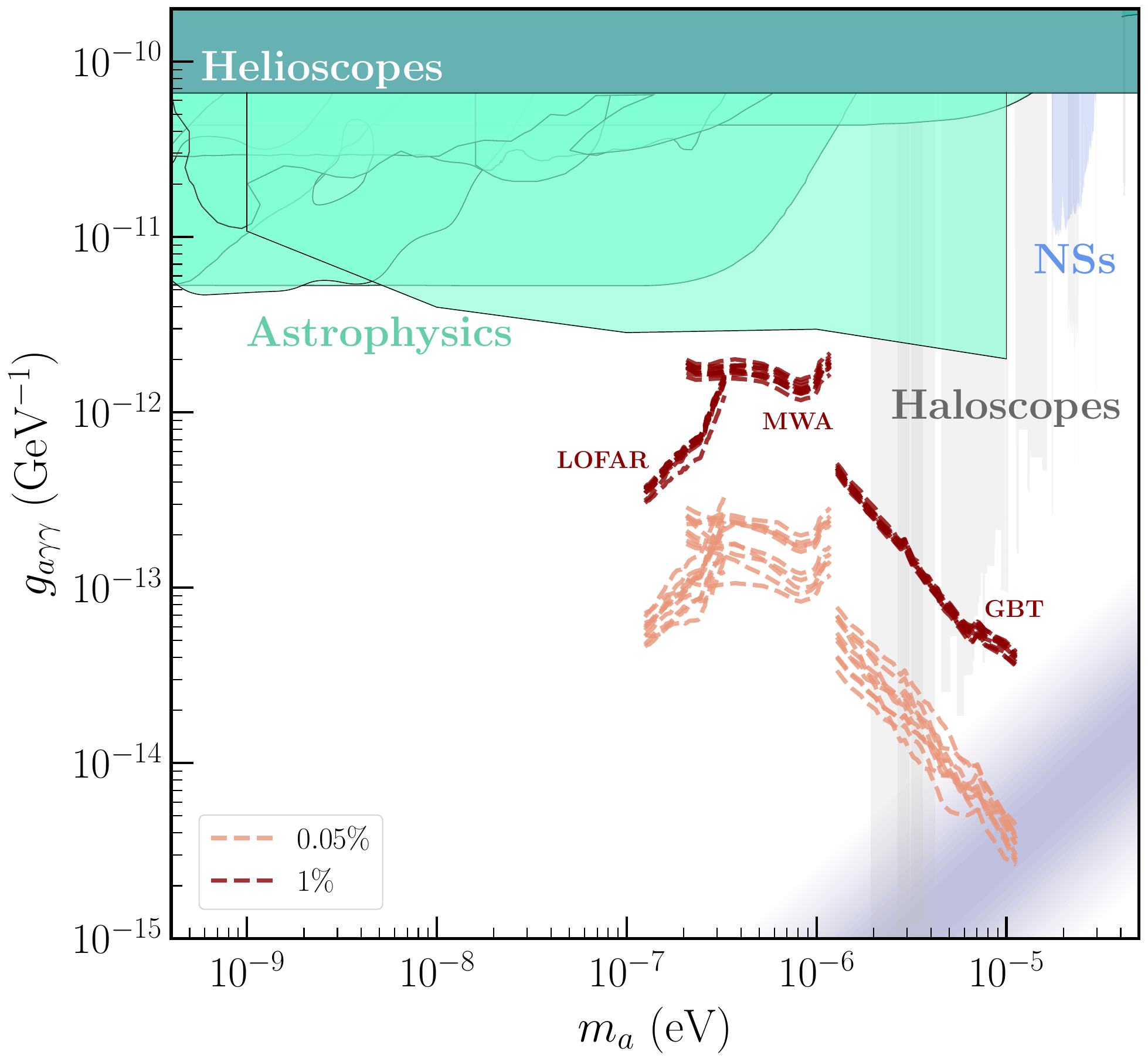}
    \includegraphics[width=0.49\textwidth]{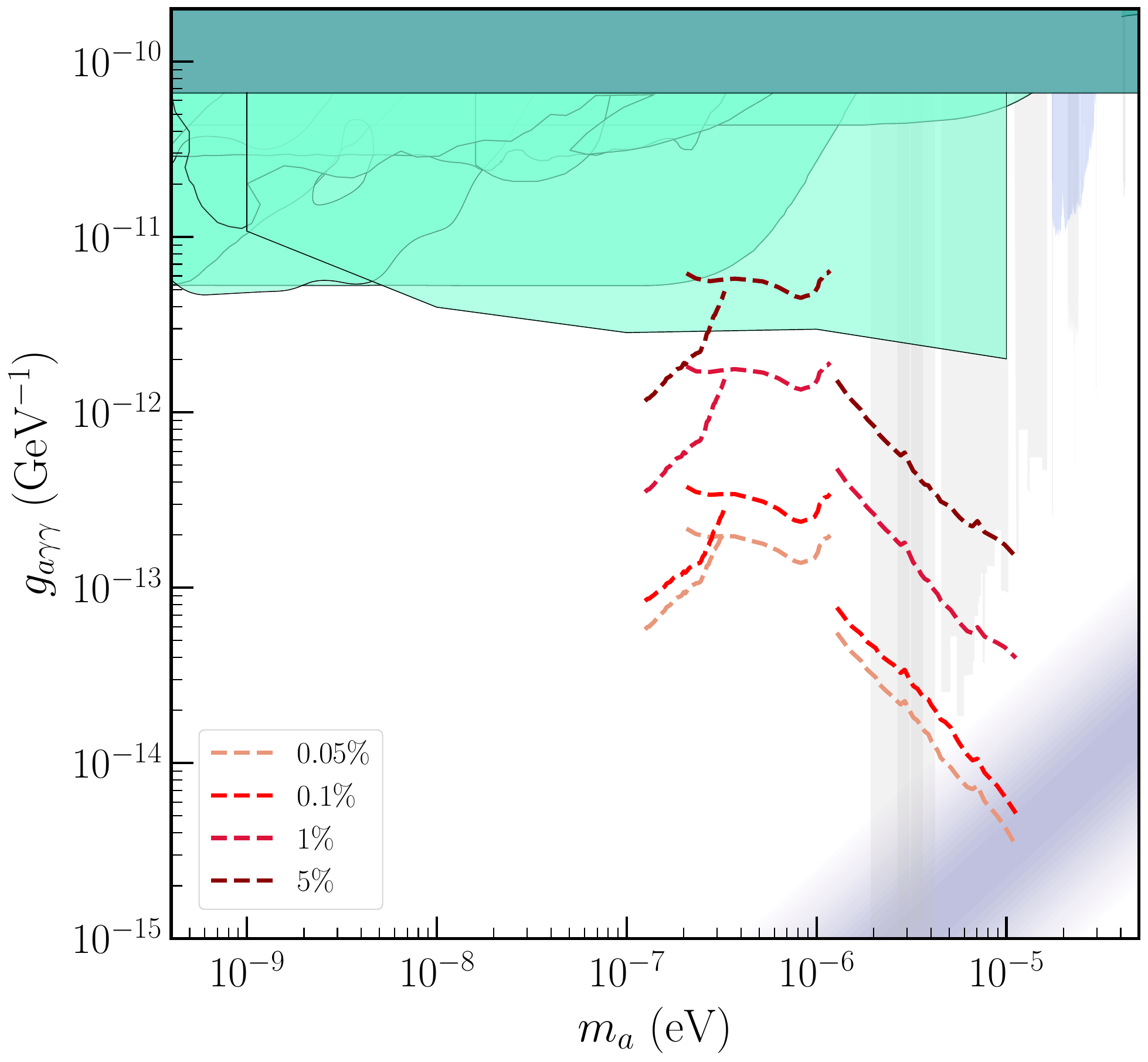}
    \caption{Projected sensitivity to the radio end-point arising due to resonant photon production from axion bound states -- sensitivity estimates are shown for LOFAR (left-hand set of curves), MWA (central set of curves), and GBT (right-hand set of curves). {\it Left:} We plot the projections from ten different realizations of our population simulation, requiring either $0.05 \%$ or $1 \%$ of the neutron star population to have flux density above threshold. {\it Right:} We instead plot the average projected sensitivity across the ten populations, further varying the required number of observable neutron stars. Our projections are compared to current constraints from neutron stars~\cite{Foster:2022fxn} (shown in blue), haloscopes~\cite{Sikivie1983,DePanfilis:1987,Hagmann:1990,Hagmann:1998cb,Asztalos:2001tf,Asztalos:2009yp,Du:2018uak,Braine2020,Bradley2003,Bradley2004,Shokair2014,HAYSTAC,Zhong2018,Backes_2021,mcallister2017organ,QUAX:2020adt,Choi_2021,Alvarez_Melcon_2021} (gray), helioscopes~\cite{Anastassopoulos2017} (blue green), and astrophysics~\cite{Wouters:2013hua,HESS:2013udx,Payez_2015,Fermi-LAT:2016nkz,Meyer:2016wrm,Marsh:2017yvc,Reynolds:2019uqt,Xiao:2020pra,Li:2020pcn,Dessert:2020lil,Dessert1_2022,Dessert2_2022,Noordhuis:2022ljw} (light green). The first two of these categories are drawn with reduced opacity to highlight that they require axions to be dark matter. The QCD axion band is shown in purple~\cite{DiLuzio:2020wdo}.}
    \label{fig:Limits}
\end{figure*}

We use Eqs.~\ref{eqn:telescope_sensitivity} and~\ref{eq:array} to compute the minimum observable flux densities for both continuous and transient emission. We emphasize that we use the term `transient' to describe radio emission from axion cloud decay at neutron star death, although the duration may be considerably longer than the total observing time of the radio emission (see Fig. \ref{fig:TransTime}). When considering the minimum flux density for a transient, we take $\tau = {\rm Min}\left[\tau_{\rm dur}, \tau_{\rm obs}\right]$, where $\tau_{\rm dur}$ is the transient duration and $\tau_{\rm obs}$ is the total observation time. 

We estimate the sensitivity using the Low-Frequency Array (LOFAR; sensitive to axions in the mass range $\sim (1.2 - 3.3) \times 10^{-7} \, \rm eV$), the Murchison Widefield Array (MWA; for axions in the mass range $\sim (2.6 - 12) \times 10^{-7} \, \rm eV$), and the Green Bank Telescope (GBT; for axions in the mass range $\sim (1.2 - 10) \times 10^{-6} \, \rm eV$). Note that for GBT, we only take into account sensitivity at the lower part of the observable frequency range, as we do not consider axion masses $m_a > 10^{-5} \, \rm eV$. For each telescope, the minimum detectable flux depends on the combination $T_{\rm syst}/G$, which is called the system equivalent flux density (SEFD). In the relevant frequency ranges mentioned above, the SEFDs were determined to be in the range $\sim (30 - 60) \, \rm kJy$ for LOFAR~\cite{LOFAR2013}, $\sim (70 - 750) \, \rm kJy$ for MWA~\cite{Sutinjo_022}, and $\sim (10 - 50) \, \rm Jy$ for GBT~\cite{GBTSense}\footnote{The GBT sensitivity can also be determined using the online sensitivity calculator, found \href{https://dss.gb.nrao.edu/calculator-ui/war/Calculator_ui.html}{here}}. LOFAR is composed of 52 stations, MWA of 128 observing elements, and GBT is a 100-m single dish telescope. Taking $N_{\rm pol} = 2$, fixing the observing time to be $1$ hour, and setting the bandwidth to be $0.05 \times m_a$ (roughly the maximal width of the features of interest), the requirement of an SNR level of 5 corresponds to minimum observable flux densities in the range $\approx (25 - 50) \, {\rm mJy}$ for LOFAR, $\approx (6 - 310) \, {\rm mJy}$ for MWA, and $\approx (47 - 720) \, \mu{\rm Jy}$ for GBT. We use the frequency-dependent $S_{\rm min}$ determined above to estimate the sensitivity to continuous and transient emission in Figs.~\ref{fig:Limits} and~\ref{fig:TransObs}, respectively (except in the cases where the transient duration is shorter than an hour; here we compute with a smaller observing time).

\subsection{Kinematic end-point}
First, we turn our attention to estimating the sensitivity of existing radio telescopes to the radio end-point shown in Fig.~\ref{fig:Spectra2}, which we achieve by simulating the population of neutron stars within our Milky Way galaxy. 

Neutron stars form from the collapse of main-sequence stars with $ 8  \, M_\odot \lesssim M \lesssim 25 \, M_\odot$. The core-collapse supernova (CCSNe) rate in the galaxy is thought to be on the order of two per century -- since most CCSNe result in the formation of a neutron star (with just a small fraction going into black holes), we adopt a maximally conservative estimate of the neutron star formation rate of $\dot{N}_{\rm NS} = 1 / 100 \, \rm yrs$ (taken for simplicity to be constant over the history of the Milky Way). We further assume that the spatial distribution of neutron stars follows the galactic disk (which we take to extend out to a distance of $10 \, \rm kpc$), and that any neutron star older than $10^{6} \, \rm yrs$ has crossed the death line (and is thus no longer producing continuous emission). This generates a population of $\sim 10^4$ active neutron stars with ages $\tau_1, \cdots, \tau_N$. The properties of each neutron star in the simulated population are taken to be one of the sample neutron stars listed in Table~\ref{tab:pulsars} (the assignment being done at random), but evaluated at the appropriate age $\tau_i$. 

It is worth noting that we adopt a very conservative neutron star population model. To date, more than 3600 pulsars have been observed~\cite{manchester2005australia,ATNF}, most of which are located in the local neighborhood of the galaxy. Additionally, most pulsar emission (both in the radio band and in gamma-rays) is highly beamed, suggesting the true fraction of active local neutron stars should be much higher than the observed population. Correcting for these observational biases, and dropping the assumption of a sharp truncation on the neutron star lifetime at $10^{6} \, \rm yrs$ (which is very conservative given the large number of neutron stars observed with ages of $\tau \sim \mathcal{O}(10) \, \rm Myr$, see \eg~\cite{igoshev2019ages}), typically yields an active neutron star population roughly 2 orders of magnitude larger~\cite{faucher2006birth}. In future work we intend to investigate the extent to which a more careful treatment of neutron star populations will allow us to extend our projected sensitivity. 

We compute the flux density in the radio end-point for each of the neutron stars in ten different realizations of the Milky Way population, and determine the coupling for which the flux of at least either $5$ or $100$ stars (corresponding to $0.05\%$ and $1\%$ of the total population respectively) exceeds the flux density thresholds outlined in the previous section\footnote{In general, the observation of a single spectral line could be sufficient in order to detect axions. We adopt more conservative thresholds for two reasons: $(i)$ our population modeling is more uncertain in the tails of the neutron star distribution, and thus adopting different thresholds indirectly reflects the sensitivity to this aspect of the analysis, and $(ii)$ the joint detection of this feature in multiple neutron stars across the population would provide striking evidence for the existence of axions.}. Note that we compute the flux densities only at frequencies corresponding to the points $m_a = 10^{-7} \, \rm eV$, $m_a = 10^{-6} \, \rm eV$, and $m_a = 10^{-5} \, \rm eV$, interpolating these results to estimate the flux densities at intermediate frequencies. This is done in order to maintain computational tractability, given that the flux density calculations require a significant amount of CPU hours. The projections determined using this procedure are shown in the left-hand panel of Fig.~\ref{fig:Limits}, where each line corresponds to a different realization of the neutron star population. In the right-hand panel of Fig.~\ref{fig:Limits} we further illustrate the change in sensitivity as we shift the minimum number of observable neutron stars, taking on values between $0.05 \%$ and $5 \%$ of the total population. The lines shown in this panel reflect the average sensitivity realized over the ten simulated populations. In all simulations we find that the neutron stars with the largest flux densities have distances neither particularly close nor particularly far from Earth. The populations are furthermore not dominated by any single neutron star in our sample (rather, five of the neutron stars tend to produce overall stronger fluxes than the others). We do note, however, that the best projections are generally set by younger neutron stars, having ages $1 \, \textrm{kyr} \lesssim \tau_i \lesssim 10 \, \textrm{kyrs}$. At larger couplings, when neutron stars are more abundant, older neutron stars that are closer to Earth can instead dominate the signal.

Fig.~\ref{fig:Limits} shows that the continuum radio emission for axion masses $m_a \sim 10^{-5} \, {\rm eV}$ has the ability to probe the QCD axion band -- we expect this trend to extend to slightly higher masses, potentially as high as $m_a \sim 10^{-4} \, \rm eV$. Nevertheless, we have chosen not to present results for the entirety of the high mass regime as the modeling of the axion production rate becomes less reliable. This is a consequence of the fact that high-energy axions are sourced from very small-scale fluctuations in the electromagnetic field, and these appear only in the highly non-linear phase of the damping process. In the future we hope to improve the modeling in this regime so that more reliable predictions at large axion masses can be made.

\begin{figure*}
    \centering
    \includegraphics[width=0.96\textwidth]{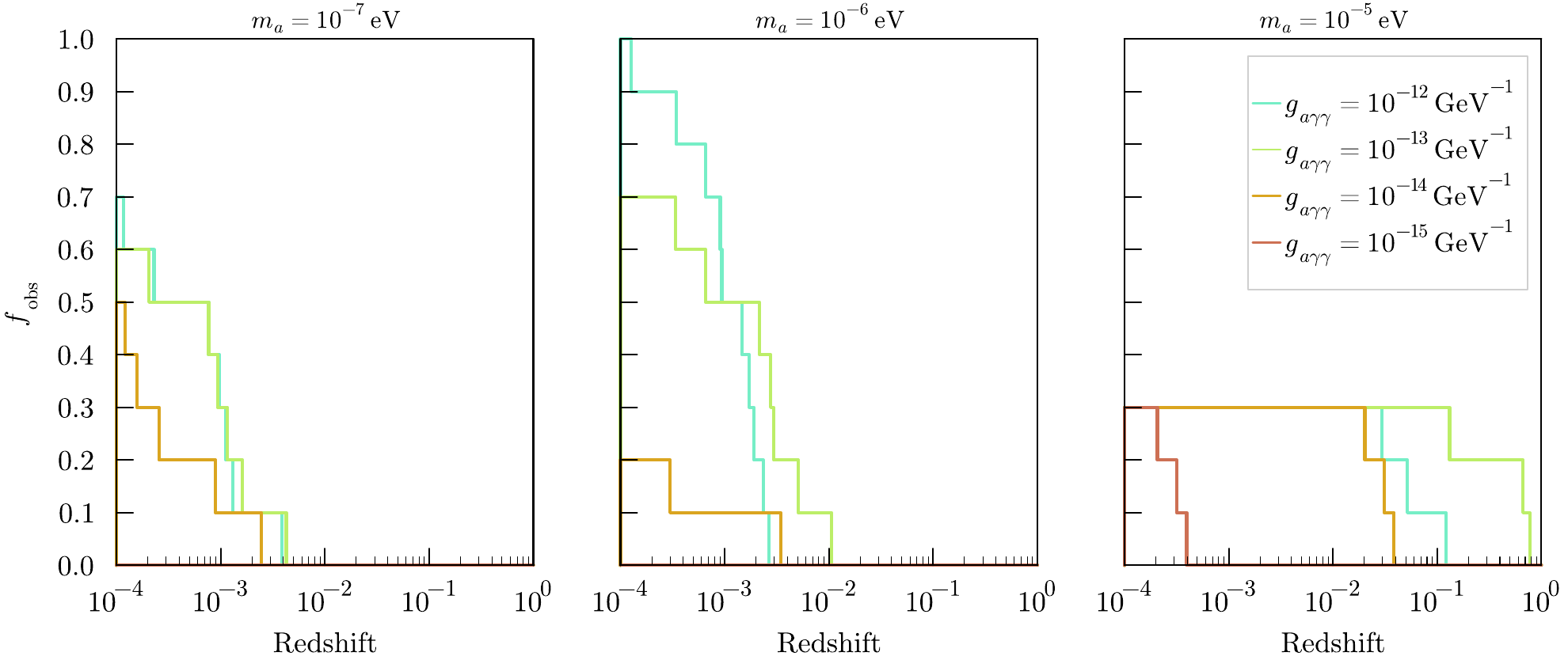}

    \caption{Fraction of observable transient events as a function of redshift and axion-photon coupling, plotted at three different values of the axion mass. At $m_a = 10^{-5} \eV$ we do not find values for $f_{\rm obs}$ above $0.3$, since at this mass only three of our neutron star samples produce a transient. We assume detection thresholds as calculated at the start of Sec.~\ref{sec:sensitivity}. }
    \label{fig:ObsHist}
\end{figure*}

\subsection{Transient decay}\label{sec:transient_rate}
In this section we assess the observability of transient events, working under the assumption that each neutron star undergoes a single transient decay which occurs instantly when crossing the death line. This is likely an oversimplification. It is unclear how quickly neutron stars will evolve from the active to the dead state, and also whether transients can be induced earlier in a neutron star's lifetime from temporary rearrangements of charges in the magnetosphere. Such short-term changes in the magnetospheres of young pulsars can, for example, arise when the star experiences a glitch, \ie a sudden spin-up that is expected to occur when angular momentum is exchanged between the superfluid interior and the solid crust, see \eg~\cite{philippov2022pulsar}\footnote{It is worth highlighting that glitches are expected to induce sizable shifts in the resonant conversion surface on timescales comparable to the light crossing time. Such a shift could open resonant conversions in parts of the cloud which had previously been kinematically blocked, leading to a rapid dissipation of radio energy on $\mathcal{O}({\rm ms})$ timescales. It is conceivable that such rapid bursts can contribute to the observable population of fast radio bursts; we leave a more detailed investigation of this phenomenon to future work.}. As a result, the sensitivity to the transient decay feature is expected to be notably more uncertain than that of the kinematic end-point. Nevertheless, given that the energy near the neutron star surface must be dissipated, we expect that the calculations performed here still provide reasonable benchmark estimates that also hold in more general scenarios.  

In order to estimate the observability of transient decays one must know the transient event rate, since this effectively defines the typical distances of transients from Earth. The lifetime of progenitor main sequence stars, $\tau_{\rm CC} \approx 55 \, \textrm{Myr} \, (M / 8 \, M_\odot)^{-2.5}$, and the characteristic timescale required for neutron stars to cross the death line, $\tau_{\rm NS} \approx 10 \, \rm Myr$, are negligible on cosmological timescales. As a result, we can assume the transient rate follows  the cosmological CCSNe rate, $R_{\rm CC}(z) = \xi \psi(z)$. Here $\psi(z)$ is the star formation rate and $\xi = 6.8 \times 10^{-3} \, M_\odot^{-1}$ is the efficiency to form CCSNe, which we compute using the Salpeter initial mass function. For $\psi(z)$ we adopt the best-fit function to UV and IR galaxy surveys derived in~\cite{Madau2014}, giving us a CCSNe rate of
\begin{align}\label{eq:rcc}
    R_{\rm CC}(z) \simeq 10^{-4} \, {\rm yr}^{-1} \, {\rm Mpc}^{-3} \frac{(1 + z)^{2.7}}{1 + [(1 + z)/2.9]^{5.6}} \, .
\end{align}

Only a fraction, $f_{\rm obs}$, of transients will be observable from redshift $z$. Here, we estimate this as the fraction of our ten sampled pulsars that would have a peak transient flux density above the thresholds defined at the start of this section. We plot $f_{\rm obs}$ as a function of redshift and axion mass in Fig.~\ref{fig:ObsHist} for various axion-photon couplings; here we have neglected the impact of redshifting of the central frequency, as for the couplings of interest observable transients tend to be concentrated toward low redshifts where this effect is negligible. One can observe a similar scaling with coupling in $f_{\rm obs}$ as was seen in Fig.~\ref{fig:FluxEvol_1NS} -- notice hereby that at $m_a = 10^5 \, \rm eV$ conversions are closest to the neutron star and thereby most adiabatic, leading to a relatively sharp decrease in $f_{\rm obs}$ when one moves to large couplings. It is worth mentioning that our estimates of $f_{\rm obs}$ are based on a small population of only ten samples -- it is possible that a small number of rare events (with $f_{\rm obs} \ll 10\%$) actually dominate the total event rate, since such events could be observable to much larger distances. The computational complexity of evaluating transient events for large populations of neutron stars unfortunately makes estimating the tails of this distribution currently unfeasible, and so we defer a more detailed study of a larger population to future work.

\begin{figure*}
    \centering
    \includegraphics[width=0.96\textwidth]{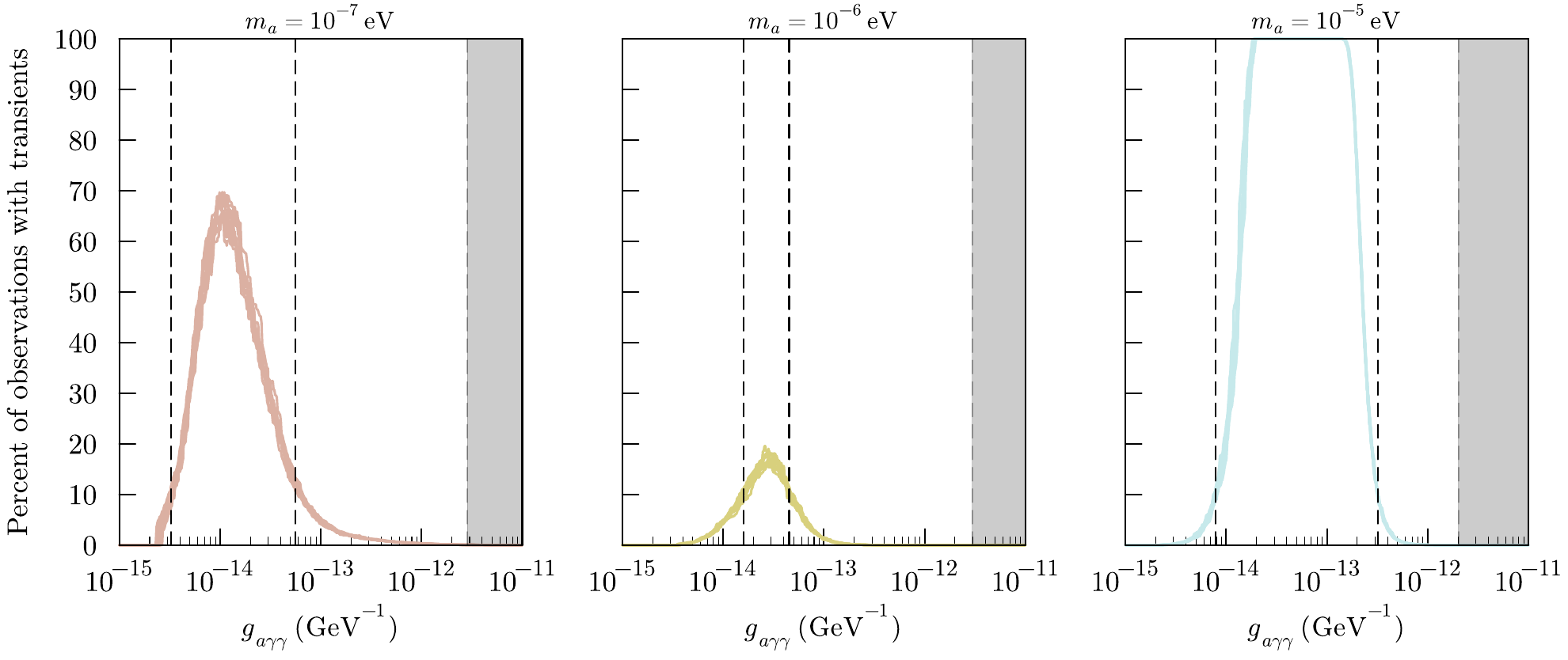}

    \caption{Percentage of observations within our observation scheme that contain at least one transient event. Results are shown for three fixed axion masses as a function of axion-photon coupling. The thickness of the bands reflects the impact of generating ten different realizations of the observing procedure. The vertical dashed lines highlight the coupling at which $10 \%$ of observations contain a transient. Previously excluded regions of parameter space~\cite{Noordhuis:2022ljw} are shaded in gray. }
    \label{fig:TransObs}
\end{figure*}

\begin{figure}
    \includegraphics[width=0.48\textwidth]{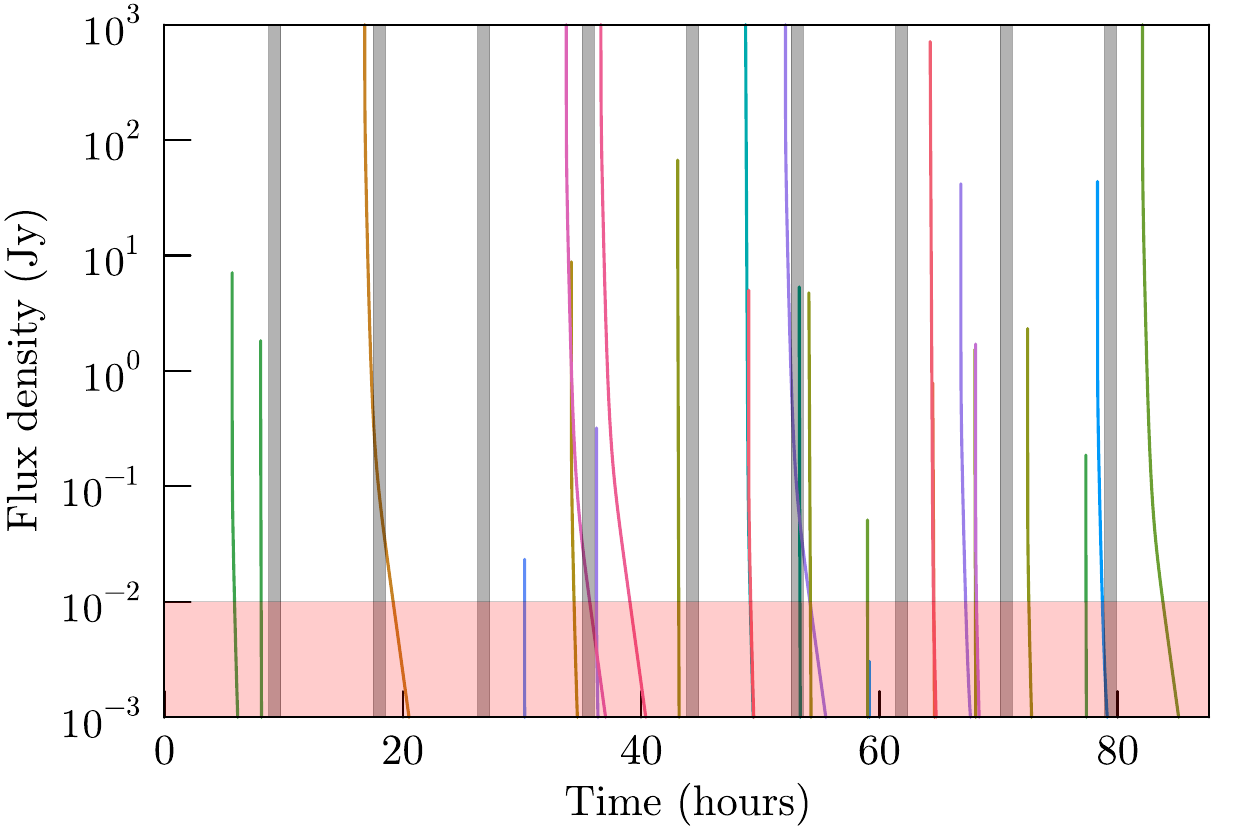}
    \caption{Schematic illustration of the observational procedure used to generate the results in Fig.~\ref{fig:TransObs}. Here, transient events are randomly distributed over the observing window $\tau_{\rm window}$. The gray vertical bands illustrate the characteristic observing times (for the purpose of illustration, we have chosen to plot 1$\%$ of total observations, and have taken these to be regularly spaced), and the flux threshold is highlighted in  red; observable transients are thus identified as those which cross the gray bands while remaining above the red band (as mentioned earlier, for transients shorter than the observing time the flux threshold is actually higher -- this is taken into account, but not indicated in the plot). This illustration has been made using an increased observable transient rate for clarity. }
    \label{fig:TransExample}
\end{figure}

The observable transient rate can be obtained by combining Eq.~\ref{eq:rcc} with the results shown in Fig.~\ref{fig:ObsHist}, and is given by
\begin{align}\label{eqn:transient_rate}
    \dot{N}_{\rm obs}(g_{a\gamma\gamma}) \simeq \int dV_c \, f_{\rm obs}(z, g_{a\gamma\gamma}) R_{\rm CC}(z) \, ,
\end{align}
where $dV_c =(1+z)^2 D_A(z)^2 / H(z) \, d\Omega dz$ is the comoving volume element, $D_A(z)$ is the angular diameter distance, and $H(z)$ is the Hubble parameter. Eq.~\ref{eqn:transient_rate} describes the rate at which transient decays are observable across the full sky -- radio telescopes, however, observe only a fraction of the sky, and only for a fraction of the day. In order to establish a notion of transient observability based off of Eq.~\ref{eqn:transient_rate}, we therefore devise a demonstrative observation scheme. Namely, we: $(i)$ choose a sufficiently large observing window (so that for each axion mass and axion-photon coupling $\tau_{\rm window} \gg \dot{N}_{\rm obs}^{-1}, \, \tau_{\rm dur}$), $(ii)$ populate this window randomly with $\left< N\right> = \tau_{\rm window} \times \dot{N}_{\rm obs}$ transient events, $(iii)$ sample $10^3$ random observation times within the observation window (each assumed to have a duration of 1 hour), and finally $(iv)$ record the percentage of these observations that contain at least one observable transient event. 

We perform ten realizations of the above procedure at a range of axion-photon couplings for axion masses $10^{-7} - 10^{-5} \eV$. The results are plotted in Fig.~\ref{fig:TransObs}. The observation scheme itself is furthermore illustrated in Fig.~\ref{fig:TransExample} for a single choice of parameters. Interestingly, we find that there is an optimal range of couplings that provides the highest transient observability. This follows partly from the earlier arguments made regarding the scaling of the transient flux density with axion-photon coupling (see the discussion concerning Figs.~\ref{fig:FluxEvol_1NS} and~\ref{fig:ObsHist}). The decrease in the percentage of observable transients that can be seen at high couplings is a consequence of shorter transient duration.

The procedure outlined above does not account for the fact that radio telescopes view only a small fraction of the sky (for example, LOFAR has a full width at half maximum on the order of a few degrees). In order to correct for this we note that typical telescopes observe $\sim 5000$ hours per year, with many having been in operation for more than a decade. Assuming for a moment that there exists one observable event in every hour of observation, one would expect LOFAR to have observed on the order of $( 0.006 \, {\rm str} / 4\pi \, {\rm str}) \, \times \, 1 \, {\rm event / hour} \, \times \, 5000 \, {\rm hours / year} \, \times \, 16 \, {\rm years} \sim 38$ events. At small and large couplings, however, the fraction of observable events per one hour of observing time is significantly reduced (see Fig.~\ref{fig:TransObs}). We choose to estimate the detection thresholds as the couplings for which $\mathcal{O}(10\%)$ of observations have a transient present in the sky -- in the case of LOFAR, this would correspond to roughly $\sim 4$ observed events. Our detection thresholds are highlighted in Fig.~\ref{fig:TransObs} using vertical dashed lines. 

It is worth emphasizing that the prediction of axion clouds around neutron stars is remarkably generic, relying only on the premise of the existence of an axion with a mass roughly in the window $10^{-9} \, {\rm eV} \,\lesssim m_a \lesssim 10^{-4} \, {\rm eV}$ and a coupling to electromagnetism. Note that this is far more conservative than the assumptions from related axion searches which require axions to contribute to the dark matter density in the galaxy~\cite{Pshirkov:2007st,Huang:2018lxq,Hook:2018iia,Leroy:2019ghm,Safdi:2018oeu, Battye:2019aco,Foster:2020pgt, Witte2021,battye2021robust,Foster:2022fxn, Witte:2022cjj,Battye:2023oac}, and in some cases further require axion dark matter to be predominantly confined to small-scale gravitationally bound structures~\cite{Iwazaki:2014wka,Bai:2017feq,Dietrich:2018jov,Prabhu:2020yif,Edwards:2020afl,Buckley:2020fmh,Nurmi:2021xds,Bai:2021nrs,Witte:2022cjj}. In addition, the radio signatures produced by axion clouds are easily differentiated from both astrophysical backgrounds and radio signals arising from axion dark matter -- this is due to the fact that the radio emission is confined to a frequency range $m_a \lesssim \omega \lesssim \sqrt{m_a^2 + k_{\rm esc}^2}$, and thus is far more narrow in frequency space than smooth astrophysical radio emission, and far wider than radio emission sourced by axion dark matter (which is expected to have a width $\delta \omega / \omega \sim \mathcal{O}(10^{-5})$~\cite{Witte2021}).

\subsection{Uncertainties in the magnetosphere}
The quantitative results in this work have been obtained under the assumption that the magnetosphere is composed of a dipolar magnetic field and a non-relativistic (largely charge-separated) highly magnetized plasma, and that this configuration is approximately stable on timescales shorter than the timescale of magnetorotational spin-down of the neutron star itself. While these are all reasonable first-order approximations, deviations are known to arise in certain regimes; here, we briefly comment on the extent to which such deviations can impact our projected sensitivity, deferring a more detailed numerical analysis of these effects to future work.

Let us start by addressing the question of the magnetic field structure. In general, pulsar spin-down is dominated by large-scale energy losses near the light cylinder, and thus high-precision measurements of the spin-down rate yield direct estimates of the dipolar field strength (all higher order multipoles being negligible at large radii), see \eg~\cite{SashaReview}. Near the surface, however, higher order multipoles can dominate over the dipole component; in fact, observations of millisecond pulsars appear to favor the presence of \eg quadrupole moments in some systems (although caution should be taken in extrapolating from the millisecond to the isolated pulsar population)~\cite{riley2019nicer}. Should a large quadrupole moment be present, the magnitude of both $\vec{B}$ and $\vec{E}$ in the gaps will increase, and the geometry of the open field line configurations will be modified (with one polar cap being shrunk, and the other being turned into an extended annulus around the star -- see \eg~\cite{Gralla:2017nbw} for an illustration). These effects can induce a shift in both the efficiency and spectrum of axions produced from the gap collapse process. Note that a larger $\vec{E} \cdot \vec{B}$ field enhances axion production, while the quadrupole geometry shrinks the gap volume, thereby largely offsetting any net shift in the axion production rate. While precise calculations will need to be performed in order to quantitatively assess the evolution of these systems with higher order multipoles, $\mathcal{O}(1)$ multipolar field strengths are not expected to dramatically alter the picture presented here.

The assumption of a non-relativistic, largely charge-separated plasma is expected to be a very good approximation for the near-field closed field line zones of isolated radio pulsars. This approximation can, however, break down along the open magnetic field line bundles, where pair production can source an inhomogeneous relativistic pair plasma that streams out away from the star. Kinematics constrain the maximal charge multiplicity, defined as $\lambda \equiv n_{\pm} / n_{\rm GJ}$, to values $\lambda \lesssim \mathcal{O}(10^5)$~\cite{Timokhin:2018vdn}, with simulations suggesting that the typical value across open field line bundles is closer to $\lambda \sim 10^3$~\cite{Timokhin:2012sk,Cruz:2021hku}. The characteristic Lorentz factor of the bulk of the plasma is typically $\left< \gamma\right> \sim \mathcal{O}(10)$~\cite{Chen:2019wwr}, suggesting that the effective plasma frequency in the open field line bundle differs by a factor $\propto \sqrt{\lambda} / \left< \gamma \right>^{3/2} \sim \mathcal{O}(1)$. Since the open field line regions are volumetrically small (for radii $r \lesssim \mathcal{O}(10^2) \, \rm km$), and the deviation is not dramatic, a more detailed treatment of the inhomogeneous plasma distribution in the open field line regions is not expected to significantly alter our results.

Finally, there are two other scenarios that could lead to non-trivial modifications of the evolution of axion clouds that are worth commenting on. The first relates to the appearance of pulsar glitches. Here, the sudden shift in the rotational period (and potentially in the magnetic field strength) serves to abruptly modify the axion production rate and the location of the conversion surface. The axion cloud will attempt to re-equilibrate, a process which typically takes place on short timescales relative to the age of the neutron star -- as such, the out-of-equilibrium phase of the system, in which our approximations break down, represents only a minor deviation from our fiducial evolutionary model. A second concern arises in the case where the neutron star is undergoing significant accretion (\eg from a binary companion). Should a significant amount of matter get funneled down the open field line regions, the matter could serve to screen the electric field in the vacuum gaps, suppressing axion production and altering the growth history of the cloud. The binary formation timescale, however, tends to be large, and thus this effect is expected to prohibit axion cloud formation only around very old neutron stars -- these stars are not ideal candidates for detection. Moreover, one can use the existence of radio emission from the open field line regions itself as an indication of whether a system is in such a state (meaning for a given system one is not entirely ignorant of accretion playing a role in the evolution of the cloud). Future work addressing systematic uncertainties from population synthesis models for axion cloud evolution will need to address the extent to which these effects can inhibit (or potentially enhance) detection prospects.

\section{Conclusions}\label{sec:conclusions}
In this paper we have shown that if there exists an axion with mass in the range $10^{-9} \, {\rm eV} \lesssim m_a \lesssim 10^{-4} \, \rm eV$ that couples to electromagnetism, then all active neutron stars are expected to be surrounded by dense axion clouds. This is an inescapable consequence of the fact that non-relativistic axions can be copiously produced in the polar caps of neutron stars and gravitationally bound to the star. Owing to their feeble interactions, these bound state axions struggle to dissipate their energy, leading to substantial growth of the local axion density on astrophysical timescales. 

Using a synthesized population of neutron stars, we have shown that the typical densities of axion clouds near the surface of the neutron star can easily reach and exceed $10^{22} \, {\rm GeV \, cm^{-3}}$, generating environments in which the large axion number densities can compensate for the feeble nature of their interactions with the Standard Model. Importantly, the quoted densities at high couplings, $g_{a\gamma\gamma} \gtrsim 10^{-14} \, {\rm GeV^{-1}}$, have been conservatively truncated when the backreaction of the axion on the electrodynamics becomes relevant. While this is expected to be conservative, a more detailed numerical treatment which self-consistently includes the role of axions alongside the evolution of the plasma dynamics (which could be accomplished \eg by including axions in the kinetic plasma simulations of~\cite{Philippov:2020jxu,Cruz:2021hku}) will be necessary in order to better understand the properties of axion clouds in this region of parameter space.

We have also demonstrated that axion clouds will, for most axion masses and axion-photon couplings, generate radio emission via resonant axion-photon mixing. Like the cloud itself, this radio emission evolves through four phases over the course of the neutron star lifetime, generating multiple distinct signatures. These notably include a sharp end-point in the radio spectrum (roughly centered about the axion mass and with a width on the order of a few percent), and transient lines generated from the reconfiguration of charges in the magnetosphere (these can occur either during a pulsar glitch or at the end of a neutron star's lifetime when charges in the magnetosphere separate). Importantly, these signatures are not only unique, but also arise rather generically, with no underlying dependence on the distribution of axions in the Universe.

The existence of axion clouds establishes a number of new phenomenological probes for axions in the mass range $10^{-9} \, {\rm eV} \lesssim m_a \lesssim 10^{-4} \, \rm eV$, all with strong discovery potential, motivating further investigation into these environments. A full understanding will require complementary efforts from multiple branches of physics, including particle (astro)physics, plasma physics, and observational radio astronomy. This work thereby opens up a novel, cross-disciplinary field, with the capability of confirming, or constraining, the existence of one of the best-motivated candidates for new fundamental physics.

\section{Acknowledgments}
The authors would like to thank Ken Van Tilburg, Andrea Caputo, and Georg Raffelt for their valuable discussions. DN and CW are supported by the European Research Council (ERC) under the European Union's Horizon 2020 research and innovation programme (Grant agreement No. 864035 - Undark). AP acknowledges support from the Princeton Center for Theoretical Science. SJW acknowledges support through the program Ram\'{o}n y Cajal (RYC2021-030893-I) of the Spanish Ministry of Science and Innovation, through the European Research Council (ERC) under the European Union's Horizon 2020 research and innovation programme (Grant agreement No. 864035 Undark) and the Netherlands eScience Center, grant number ETEC.2019.018, and from a Royal Society University Research Fellowship (URF-R1-231065).  This article/publication is based upon work from COST Action COSMIC WISPers CA21106, supported by COST (European Cooperation in Science and Technology). We acknowledge the use of~\cite{AxionLimits} in creating the figure containing axion constraints.

\bibliography{Axion_BS.bib}

\end{document}